\documentclass[10pt,journal]{IEEEtran}

\usepackage{cite}
\usepackage[T1]{fontenc} % optional
\usepackage{amsmath}
\usepackage[cmintegrals]{newtxmath}
\usepackage{bm} % optiona
\usepackage{graphicx}
\usepackage{booktabs}
\usepackage{multirow}
\usepackage{tabularx}
\usepackage{xcolor}
\usepackage[normalem]{ulem}

\newif\ifpublicationaffiliation

\publicationaffiliationtrue       % Для подачи в журнал
\title{
Unmasking Hidden Trapping States in Breathing 
Nanopores via Structure-Informed Bayesian Inference
\\
Structure-Informed Bayesian Inference of Anomalous Transport 
and Hidden Molecular Trapping in Amorphous Media
}

\newif\ifpublicationaffiliation

 \publicationaffiliationfalse    % Для arXiv

\title{

\ifpublicationaffiliation
Unmasking Hidden Trapping States in Breathing Nanopores
via Structure-Informed Bayesian Inference
\else
Structure-Informed Bayesian Inference of Anomalous Transport
and Hidden Molecular Trapping in Amorphous Media

\fi
}

\ifpublicationaffiliation

\author{
\IEEEauthorblockN{
Andrey Ananev,
Maria Potapova,
Nikolay Kondratyuk,
Timur Vostroknutov,
Aleksey Khlyupin
}
\IEEEauthorblockA{
Moscow Center for Advanced Studies\\
20, Kulakova Str., Moscow, Russia\\
Email: andrey.ananev@phystech.edu
}
}

\else

\author{
\IEEEauthorblockN{
Andrey Ananev\IEEEauthorrefmark{1},
Maria Potapova\IEEEauthorrefmark{2},
Nikolay Kondratyuk\IEEEauthorrefmark{2},
Timur Vostroknutov\IEEEauthorrefmark{1},
Aleksey Khlyupin\IEEEauthorrefmark{1}
}

\IEEEauthorblockA{
\IEEEauthorrefmark{1}
Laboratory for Disordered Systems,\\
Phystech School of Applied Mathematics and Computer Science,\\
Moscow Institute of Physics and Technology,
Institutsky Lane 9,
Dolgoprudny,
Moscow Region,
141700,
Russia
}

\IEEEauthorblockA{
\IEEEauthorrefmark{2}
Center for Computational Physics,\\
Landau School for Physics and Research,\\
Moscow Institute of Physics and Technology
Institutsky Lane 9,
Dolgoprudny,
Moscow Region,
141700,
Russia
}

\IEEEauthorblockA{
Email: khlyupin@phystech.edu
}
}

\fi

\begin{document}

\bstctlcite{BSTcontrol}

\maketitle

\begin{abstract}
Molecular diffusion in fluctuating amorphous and macromolecular media governs key transport processes across soft-matter physics, energy storage, and biological membranes. 
Extracting localized trapping states from single-particle tracking trajectories remains a fundamental challenge; because thermal structural breathing continuously reconfigures pore boundaries, conventional geometric algorithms suffer from severe systematic biases, erroneously merging distinct localized states during cyclic molecular returns. 
Here, we address this deadlock by shifting the paradigm from local geometric recurrence to a structure-informed Bayesian regularization. 
Leveraging discrete Morse theory, we extract the time-invariant topological skeleton of the fluctuating host matrix to construct robust, gas-specific physical priors that account for individual molecular dimensions. 
Trajectory steps are sequentially partitioned via a two-stage probabilistic refinement that dynamically adapts to the transport landscape. 
Benchmarked against a rigorous environment where synthetic particles explore the actual interconnected matrix graph, our approach eliminates systemic biases, restricting macroscopic trapping parameter deviations to just a few percent under optimal linear $O(N)$ computational scaling. 
Applied to hydrogen and methane transport within a type-I kerogen matrix, serving as a prototype for highly tortuous, flexible macromolecular networks, the method successfully decodes the hidden microscopic mechanisms of confined diffusion. 
To ensure immediate broad impact, the documented open-source code and data are made publicly available, offering an accessible strategy readily adaptable to a broad spectrum of tracking phenomena, from ion transport in battery polymers to protein trafficking within cellular environments.
\end{abstract}


\section{Introduction}
% \input{introduction.tex}
% \section{Introduction2}
Molecular transport within nanoconfined amorphous materials is a central problem spanning soft-matter physics, energy materials design, and biophysics.
Among these systems, disordered cross-linked macromolecular networks present a unique challenge: their pore architecture continuously evolves due to thermal motion.
A prominent, highly complex example of such media is kerogen, the primary organic constituent of unconventional subsurface reservoirs that serves as both the source and initial host of natural gas and liquid hydrocarbons \cite{behar1987chemical,obliger2023development}.
Its pore structure consists of a disordered, cross-linked macromolecular network where interconnected cavities and narrow channels with characteristic sub-nanometer dimensions (0.1-1 nm) control the storage, transport, and recovery of confined fluids \cite{yu2022supercritical,li2023dependence}.
The dynamic, non-static reality of these fluctuating pathways, which evolves continuously with factors like thermal maturity \cite{leyssale2023replica}, makes classical continuum transport models insufficient.
Capturing the true transport mechanisms in these dynamic environments mandates a rigorous, molecular-scale description of diffusion phenomena that can bridge local structural fluctuations with macroscopic flux predictions.

Molecular dynamics (MD) simulation has emerged as the principal tool for probing nanoscale fluid transport not only in geological matrices but across a wide variety of synthetic polymers, ionic membranes, and disordered porous solids \cite{yu2021diffusion,huang2018molecular,raza2022h2,babaei2025adsorption,dawass2023prediction,parambathu2023nanoconfinement}.
Extensive MD studies on these systems demonstrate that tight confinement, surface chemistry, and matrix flexibility exert a strong, coupled influence on both thermodynamic and kinetic properties \cite{ho2016nanostructural,bonnaud2023modeling,ariskina2024confined,potier2023argon,obliger2018nanoporosity,zhang2023sorption}.
Broadly speaking, nanoconfined fluid behavior deviates profoundly from classical Fickian laws due to the specific geometry, mechanical response, and chemical composition of the confining medium \cite{nesterova2025ionic,kazemi2024wettability}.
Recent tracking studies have firmly established that molecular transport in these fluctuating environments grows sublinearly with time, exhibiting a heavy-tailed residence time distribution where local trapping events dominate macroscopic diffusivity \cite{metzler2014anomalous,zhao2018integrated,hu2023triggering}.
Despite this progress, the overwhelming majority of MD studies characterize transport exclusively via mean-squared displacement (MSD) curves averaged over the ensemble.
This traditional averaging acts as a blurry filter: it mixes completely different microscopic movements, such as quiet bouncing inside a single pore cavity and rapid jumps between pores, into a single number.
As a result, looking only at averaged curves hides the true physical mechanisms of anomalous diffusion, making it impossible to see how molecules actually move through the structural network.

To unlock these hidden mechanics, researchers increasingly turn to single-molecule trajectory analysis, a technique that has transformed our understanding of anomalous transport across biological and soft-matter systems \cite{manzo2015review,munoz2021objective,munoz2025quantitative,martinez2023sequence}.
Accurately resolving individual molecular histories is not just a mathematical exercise; it is a strict prerequisite for building reliable, physics-informed continuum models for large-scale industrial devices.
If trapping statistics are misclassified at the micro-scale, the resulting macro-models inherit severe systematic errors.
However, a fundamental distinction must be drawn between traditional mathematical frameworks and physical reality.
Classical continuous-time random walk (CTRW) models map transport as abstract stochastic jumps across an open, boundaryless bulk space.
In real amorphous matter, molecular diffusion is explicitly constrained by a highly tortuous pore topology.
Here, transport does not sample an open void; instead, it acts as a topologically conditioned random walk on a discrete graph of localized pore cavities interconnected by narrow throat channels.
Consequently, subdiffusion is not a purely energetic anomaly, but a direct consequence of spatial confinement and network connectivity \cite{scher1975anomalous}.

Existing algorithms for trapping detection typically rely on structure-blind geometric baselines, such as the conventional pairwise distance-matrix (DM) approach \cite{lanoiselee2021detecting}.
Applying these methods to anomalous transport phenomena reveals a fundamental, dual bottleneck.
First, capturing the heavy-tailed physics of subdiffusion relies entirely on long-duration trapping events, classic statistical rare events that demand massive, multi-microsecond trajectory timelines.
Paradoxically, the very length of these timelines causes conventional matrix methods to suffer from a prohibitive quadratic computational complexity, scaling as $O(N^2)$ and forcing a compromise between statistical convergence and numerical tractability.
Second, operating exclusively on trajectory geometry without knowledge of the host material induces a severe physical blindness known as cyclic trapping.
When a molecule rapidly switches back and forth between heavily coupled, adjacent pore cavities, purely geometric recurrence algorithms systematically conflate these quick inter-pore jumps with long-time intra-pore localization.
Modern machine learning approaches, while successful in idealized benchmarks \cite{munoz2021objective,munoz2025quantitative,bae2026gradcam}, have not yet accommodated the highly transient, dynamic breathing of real amorphous structures without inducing an intractable numerical overhead.
A profound methodological deadlock emerges: trajectory segmentation must simultaneously resolve spatial cyclic entanglement and accommodate continuous structural fluctuations.

A Bayesian classification framework is uniquely suited to cut through this methodological deadlock.
By assigning each trajectory step a posterior probability of belonging to either intra-trap motion or an inter-trap transition, this strategy naturally blends a prior over the two hypotheses with a physical likelihood.
When the aggregate pore-scale geometry of the host material is known, it provides exactly this likelihood: intra-trap displacements sample the interior of a pore cavity, while inter-trap transitions correspond to passage through a narrow throat channel.
Encoding these structural distributions as likelihood functions allows the classifier to resolve ambiguous cases, such as oscillatory motion near a pore boundary, that confound purely geometric methods.

In this work, we transition from localized geometric tracking to global probabilistic regularization by developing a structure-informed Bayesian (SIB) approach.
The framework leverages discrete Morse theory combined with persistent homology to process fluctuating structural snapshots of the matrix \cite{zubov2022pore}.
While the individual extracted skeletons dynamically reconfigure over time, their aggregate geometric distributions, the pore radii and throat lengths, remain statistically stationary, serving as robust, time-invariant physical priors that account for individual molecular dimensions.
Sequential step processing bypasses the need for a global recurrence matrix, establishing a linear-scaling $O(N)$ paradigm optimized for massive molecular dynamics datasets.

Furthermore, we challenge this probabilistic strategy against a rigorous benchmark environment, the trajectory simulation model (TSM), where synthetic particles explore the actual interconnected matrix graph.
This structural grounding ensures that the synthetic trajectories preserve the complete topological complexity and spatial correlations of the genuine amorphous medium, successfully reducing macroscopic trapping parameter deviations from tens of percent down to just a few percent.

To demonstrate the physical fidelity of this approach on a highly non-ideal, realistic system, we perform extensive, high-performance molecular dynamics simulations of hydrogen ($H_{2}$) and methane ($CH_{4}$) transport within a flexible, cross-linked type-I kerogen matrix.
Generating massive, multi-microsecond atomistic ensembles allowed us to accumulate thousands of consecutive transition events, successfully overcoming tail fluctuations in the subdiffusive regime.
Crucially, our analysis unmasks a profound divergence in the localized transport mechanisms of the two gases, captured directly through their network return probabilities.
Methane operates in a heavily trapped, high-recurrency regime dictated by severe geometric confinement and strong matrix interactions, forcing the molecule to execute continuous cyclic returns between adjacent pore cavities.
In sharp contrast, the smaller and weakly interacting hydrogen molecule exhibits a highly delocalized, exploratory hopping behavior with a minimal return probability.

This comprehensive single-particle tracking dataset, decoupling genuine localization from topological recurrence, is made fully accessible alongside our open-source code to ensure broad reproducibility and immediate community driven impact.
Ultimately, resolving these fine-grained single-molecule statistics establishes a clean, mathematically rigorous bridge to macroscopic continuum transport models, advancing our capacity to predict fluid fluxes in heavily disordered amorphous matter and to analyze a broad spectrum of tracking phenomena, from ion transport in battery polymers to trafficking within cellular environments.

\section{Nanoporous Matrix Architecture and Structural Dynamics}
%%%%%%%%%%%%
% Physical-system description

\subsection{Atomistic Model of the Kerogen Network}

Kerogen is a chemically heterogeneous, cross-linked organic matrix, the atomistic structure of which depends on the biological origin and thermal history of the source material. During maturation, kerogen evolves from a more aliphatic, hydrogen-rich material toward a denser and more aromatic carbon network, a process commonly represented using the van Krevelen diagram~\cite{behar1987chemical}. Because the exact atomic structure of natural kerogen is not uniquely defined, molecular simulations rely on representative molecular models whose elemental composition, density, and morphology reproduce experimentally motivated characteristics.

\begin{figure}[h]
\centering
\includegraphics[width=0.4\textwidth]{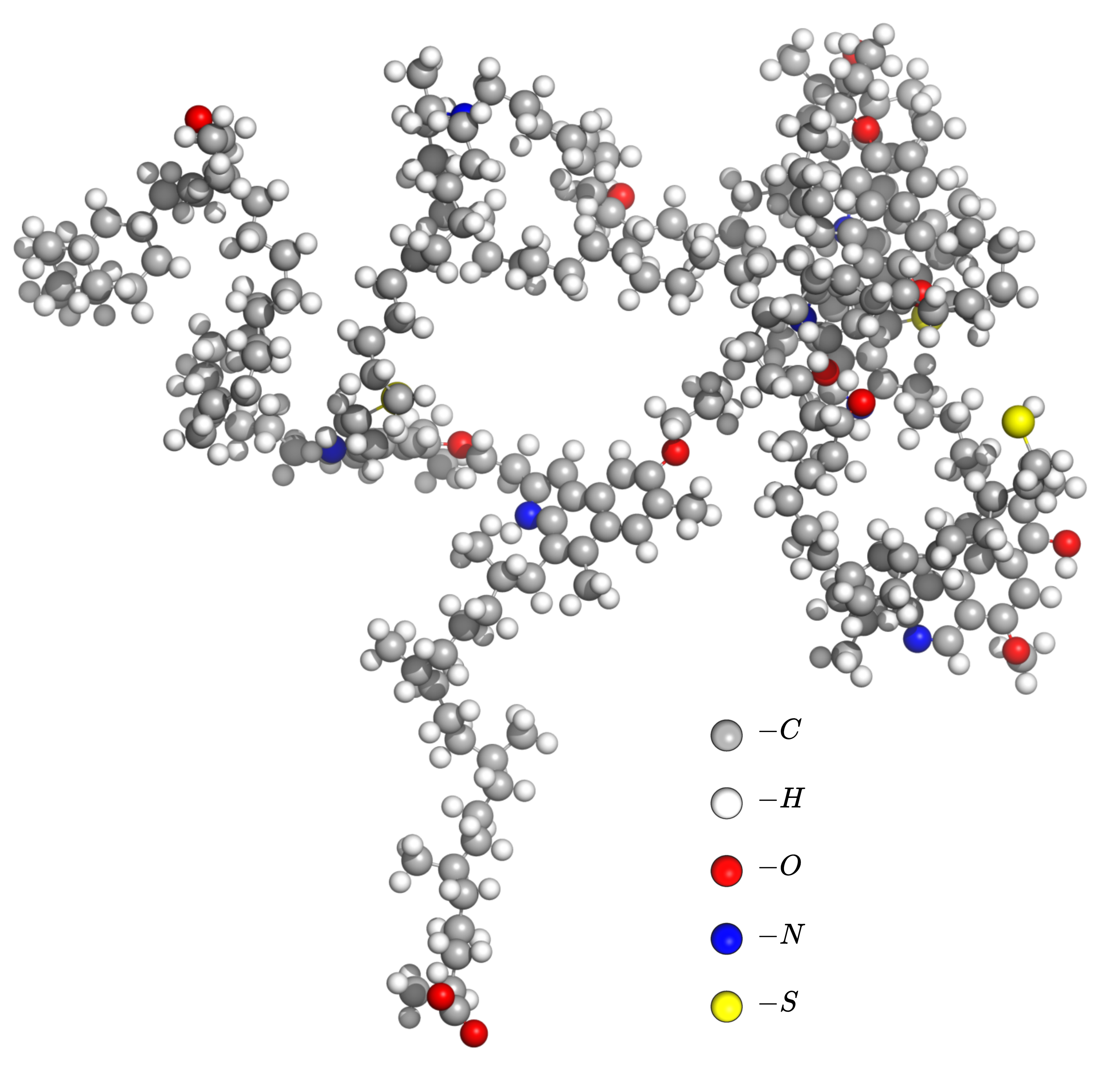}
\caption{
Two-dimensional projection of a representative kerogen macromolecule. Atom colors follow the conventional scheme: hydrogen (white), carbon (gray), oxygen (red), nitrogen (blue), and sulfur (yellow).
}
\label{fig:kerogen_molecula}
\end{figure}

In this work, we consider gas diffusion in a type I kerogen matrix. The initial kerogen molecular model follows the atomistic structures used in previous simulations of methane transport in kerogen~\cite{yu2022supercritical,yu2021diffusion}. To construct the kerogen matrix, multiple kerogen macromolecules (Fig.~\ref{fig:kerogen_molecula}) were initially placed in a large periodic simulation cell at low density and equilibrated using a simulated annealing procedure. The system was then compressed to the density of the condensed phase, and the final density was compared with experimental data for type I kerogen to validate the equilibrated structure. Further details of the matrix preparation protocol are provided in Ref.~\cite{potapova2026kerogen}. The resulting material contains a disordered void space with interconnected cavities and narrow transient channels, which provide the physical environment for methane and hydrogen transport.

\subsection{Molecular Dynamics Simulation of Nanoscale Gas Diffusion}

The initial kerogen configurations were prepared and equilibrated using  the GROMACS software~\cite{van2005gromacs,Abraham2015}. 
Kerogen intramolecular and nonbonded interactions were described with the Consistent-Valence Force Field (CVFF), which has been widely used for kerogen modeling and reproduces relevant condensed-phase properties of structures similar to kerogen~\cite{CVFF,ho2016nanostructural,huang2018molecular}. 
Methane was modeled using the TraPPE-UA parameterization adopted in Ref.~\cite{yu2022supercritical} and originally developed for transferable alkane simulations~\cite{martin1998transferable}. 
Hydrogen was modeled using the Yang--Zhong potential for hydrogen adsorption and diffusion simulations~\cite{yang2005molecular}. 
The production trajectories of methane and hydrogen analyzed in this work were subsequently generated using the OpenMM software~\cite{eastman2024openmm} to improve computational performance.

Cell equilibration was performed in the NPT ensemble at  $T=300$ K and $P=1$ bar under periodic boundary conditions. 
During this stage, temperature was controlled with the Nos\'e--Hoover thermostat~\cite{nose_molecular_1984,hoover_canonical_1985}\ with a coupling time of $0.5$ ps, and pressure was controlled with the Parrinello--Rahman barostat~\cite{parrinello1981polymorphic}. 
The production calculations were performed at $T=300$ K using the equilibrated simulation cell dimensions at fixed volume. No barostat was applied during production. 
During GROMACS equilibration, covalent bonds involving hydrogen were constrained using the LINCS algorithm~\cite{hess1997lincs}. Electrostatic interactions were computed using the particle-mesh Ewald method~\cite{essmann1995smooth}. 
The real space electrostatic and Lennard-Jones cutoffs were both set to $10$~\AA. The integration time step was $2$ fs.

The data set comprises two independent gas--kerogen simulations, one for methane ($\mathrm{CH}_4$) and one for hydrogen ($\mathrm{H}_2$). 
Each system contains $100$ kerogen macromolecules and $20$ gas molecules, yielding $20$ independent single-molecule trajectories per gas. 
To manage the massive computational output, full matrix configurations containing all kerogen atoms were saved every $250{,}000$ integration steps (frame interval of $500$~ps). 
This $500$~ps resolution is perfectly sufficient to capture the slow, thermally driven structural breathing of the macromolecular network. 
This separation of timescales ensures that we accurately track every quick, sub-nanometer transition of the gas molecules while avoiding intractable storage overhead for the rigid host matrix.
The methane simulation spans $t \approx 3.28~\mu$s and the hydrogen simulation $t \approx 3.31~\mu$s. The periodic simulation cell has dimensions $62.31 \times 74.11 \times 130.15$~\AA\ (volume $\approx 600.97$~nm$^3$).

\subsection{Thermally Driven Structural Breathing and Pore-Space Fluctuations}

\begin{figure*}[t]
\centering
\includegraphics[width=0.77\textwidth]{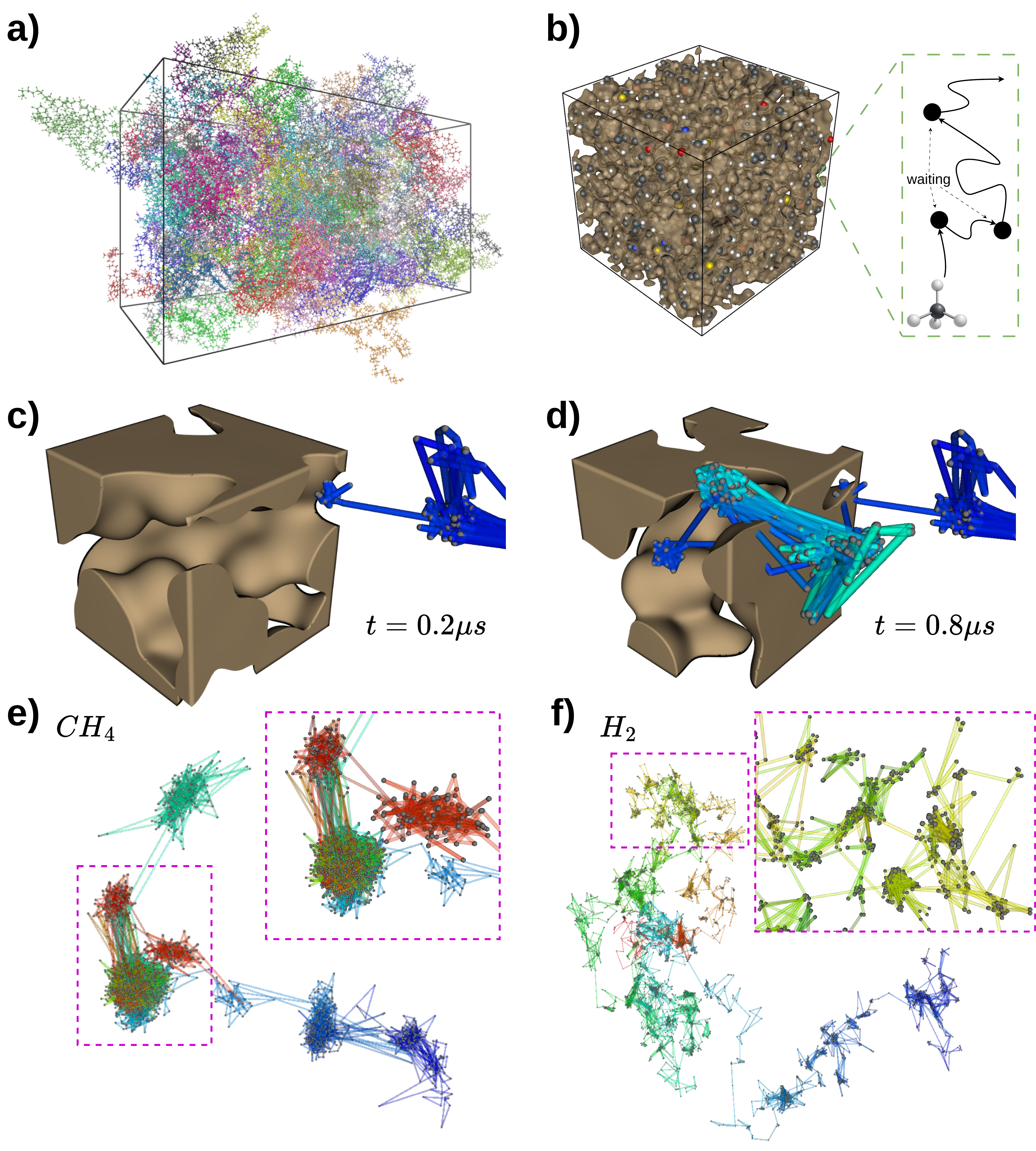}
\caption{ Multiscale representation of kerogen structure and gas transport in an evolving pore space.
(a) Simulation domain containing multiple kerogen macromolecules. Individual molecules are distinguished by color. Atomic positions, covalent bonds, and the simulation cell boundaries are shown.
(b) Representative subvolume illustrating the formation of a connected solid matrix and an interconnected pore space after merging individual molecular surfaces. 
Atoms are rendered as space-filling spheres colored by atom type (color scheme as in Fig.~\ref{fig:kerogen_molecula}). The schematic on the right illustrates gas transport as a sequence of confined motion, waiting, and rare transitions between neighboring accessible regions.
(c,d) The same kerogen subvolume with the gas molecule trajectory progressively superimposed: panel~(c) shows the path up to $t=0.2\,\mu$s, while panel~(d) extends it through $t=0.8\,\mu$s. Trajectory color encodes cumulative path length from the start of the full simulation trajectory, so the color scale is shared across both panels and the temporal progression of the molecular path is directly readable from the continuous color gradient.
(e) Full molecular dynamics trajectory of a methane molecule (CH$_4$), showing confined motion within pores and intermittent transitions between spatially separated regions.
(f) Full molecular dynamics trajectory of a hydrogen molecule (H$_2$), illustrating enhanced mobility and qualitatively different transport behavior compared with methane. 
Panels (e,f) show the respective full simulation trajectories using the same color convention.
}
\label{fig:problem}
\end{figure*}

The kerogen matrix, which serves as the primary medium for gas diffusion in this study, is structurally heterogeneous and dynamically fluctuating. In the MD simulation cell, multiple such macromolecules form a heterogeneous solid matrix with void regions that define the pore space (Fig.~\ref{fig:problem}a). At the local scale, the pore space can be represented through excluded atomic volumes and molecular surfaces (Fig.~\ref{fig:problem}b).

The kerogen cell is not static. Rather, it evolves over time due to thermal motion of its molecular components. As kerogen molecules shift positions, they intermittently open and close pathways for diffusing gas molecules. These thermally driven pore space fluctuations, referred to here as structural breathing, produce temporary confinement within pores and, conversely, transient opening of new regions for molecular diffusion.

These local fluctuations in the pore space create significant deviations from simple free diffusion. Representative instantaneous configurations of the same kerogen subvolume at two simulation times, with the same gas molecule trajectory superimposed, are shown in Figs.~\ref{fig:problem}c,d. The trajectory demonstrates confined motion within pores and rare transitions between pore regions. Apparent penetration into regions occupied by the initial solid configuration reflects temporal fluctuations of the matrix, i.e., breathing behavior, and transient opening of transport pathways.

Full molecular dynamics trajectories for methane and hydrogen are shown in Figs.~\ref{fig:problem}e and \ref{fig:problem}f, respectively. The comparison illustrates that gas transport in kerogen is dependent on the molecule type: methane exhibits more confined motion, whereas hydrogen demonstrates enhanced mobility and qualitatively different transport behavior. The dynamic nature of the kerogen structure, combined with its spatial heterogeneity, complicates the identification and classification of molecular trapping and transitions using methods based purely on trajectories.

\begin{figure}[t]
\centering
\includegraphics[width=0.45\textwidth]
 {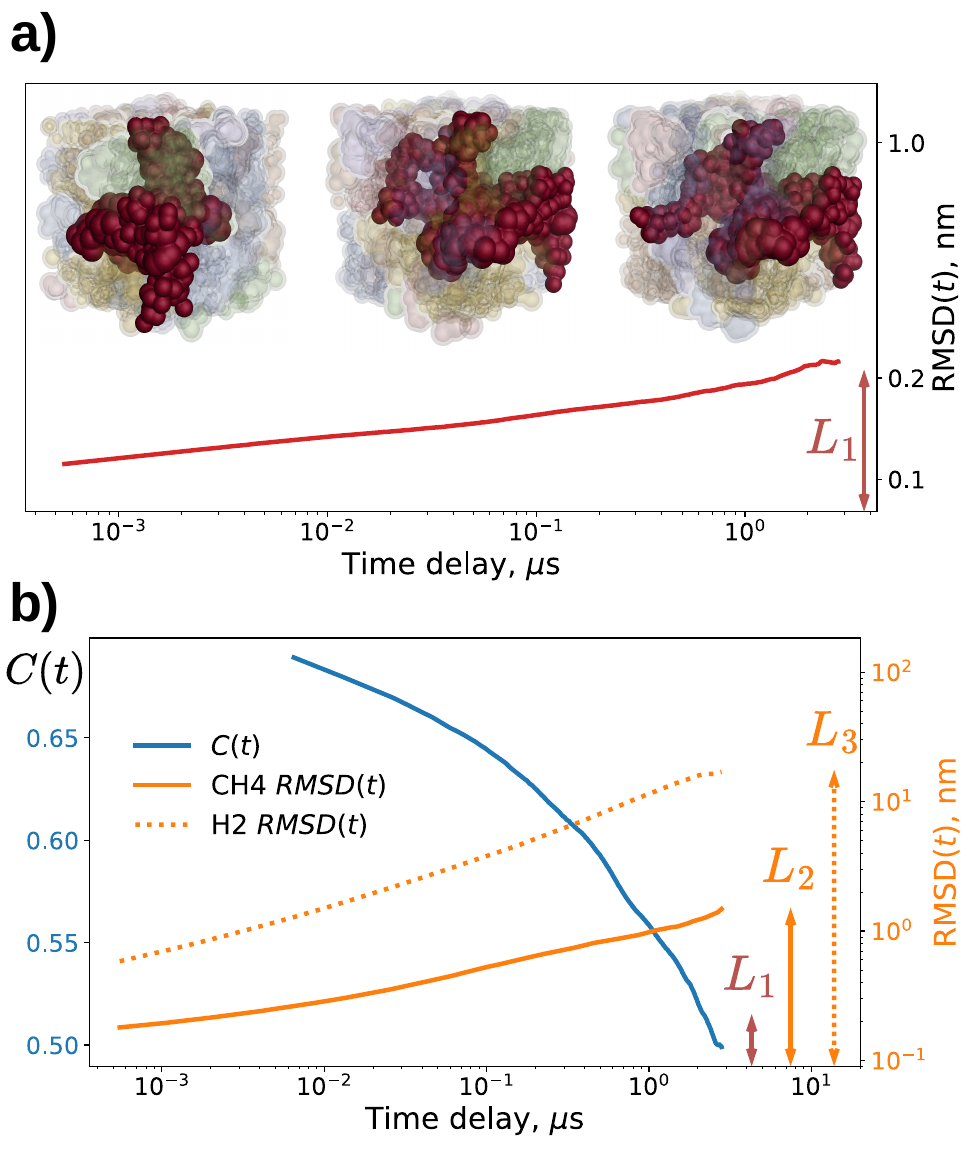}
\caption{
Relation between pore space correlations and molecular displacements in kerogen.
(a) Root-mean-square displacement $\mathrm{RMSD}(t)$ of a representative kerogen molecule, plotted in log--log scale.
The scale marker $L_1 \approx 0.2$~nm marks the total displacement accumulated by the kerogen molecule over the simulation.
The three upper subpanels show surface representations of the kerogen structure at different time instants.
One selected kerogen molecule is highlighted, while the surrounding molecules are rendered semitransparent, illustrating fluctuations of the molecule within the kerogen matrix.
(b) Autocorrelation function of the pore space, $C(t)$, shown by the blue curve, together with the root-mean-square displacements $\mathrm{RMSD}(t)$ of methane ($CH_4$) and hydrogen ($H_2$), shown by the orange solid and dotted curves, respectively.
The pore space autocorrelation is plotted against the left vertical axis, while the molecular displacements are plotted on the logarithmic right vertical axis.
Scale markers $L_1 \approx 0.2$~nm, $L_2 \approx 2$~nm, and $L_3 \approx 25$~nm are indicated on the right vertical axis of panel~(b). $L_1$ marks the total displacement of the kerogen molecule, $L_2$ the characteristic displacement of methane, and $L_3$ the correspondingly larger displacement of hydrogen over the same simulation interval.
}
\label{fig:pore_corr_and_kerogen_rmsd}
\end{figure}

To characterize the relation between matrix dynamics and gas transport, we compute the pore space autocorrelation $C(t)$ and the root-mean-square displacement $\mathrm{RMSD}(t)$ (Fig.~\ref{fig:pore_corr_and_kerogen_rmsd}).

The pore space autocorrelation $C(t) =   \frac{ \langle \psi(\mathbf{x},t) \psi(\mathbf{x},0)\rangle_{\mathbf{x}}}{\langle \psi(\mathbf{x},0)\rangle_{\mathbf{x}}}$ measures the fraction of the initial pore space that remains accessible at time $t$, where $\psi(\mathbf{x},t)\in\{0,1\}$ is the pore space indicator (unity in pore space, zero in solid kerogen). 
$C(0)=1$ by construction and $C(t)$ decreases as the spatial distribution of pore volume evolves away from its initial configuration. 

The root-mean-square displacement, $\mathrm{RMSD}(t) = \langle |\mathbf{r}(t + \tau) - \mathbf{r}(\tau)|^2 \rangle_{\tau}^{1/2}$, measures the average distance a molecule travels from its initial position, evaluated independently for the diffusing gas species and the host kerogen matrix. 
Comparing these trajectories with the relaxation of $C(t)$ highlights a pronounced separation of timescales that governs the system's transport physics. 
As visible in Figure~\ref{fig:pore_corr_and_kerogen_rmsd}b, the kerogen matrix undergoes rapid local rearrangements, with $C(t)$ exhibiting a sharp initial drop before flattening into its asymptotic bulk porosity. 
Crucially, this structural relaxation occurs orders of magnitude faster than the time required for gas molecules to diffuse across the pore network. 
By the time a methane or hydrogen molecule traverses its respective characteristic displacement lengths, $L_2 \approx 2\text{ nm}$ or $L_3 \approx 25\text{ nm}$ (indicated on the RMSD curves), the local pore environment has already fully reconfigured and entirely forgotten its initial geometric state. 
Because the local memory of the matrix is erased so rapidly compared to the macro-diffusion process, the gas molecules effectively sample a dynamically averaged, statistically invariant representation of the host medium. 
This rapid loss of correlation fundamentally justifies shifting our analytical focus from tracking instantaneous, fluctuating geometries to extracting time-invariant macroscopic statistical descriptors of the pore space. 
Thus this forms the basis of the trajectory-partitioning framework developed in the following sections.

This disparity reveals a direct physical mechanism: the pore volume surrounding the diffusing molecule is substantially redistributed on the timescale of its displacement, so the molecule encounters a continuously evolving confinement, which is the physical origin of the observed anomalous transport.

The scale markers in panel~(b) make this quantitative: $L_1 \approx 0.2$~nm marks the total kerogen molecule RMSD over the simulation, $L_2 \approx 2$~nm the characteristic displacement of methane, and $L_3 \approx 25$~nm the correspondingly larger displacement of hydrogen. The tenfold difference between $L_1$ and $L_2$ shows that the gas molecule displacements are much larger than the RMSD of the representative kerogen molecule, which is averaged over all atoms in that molecule. Nevertheless, molecular motions on the scale of $L_1$ throughout the matrix collectively produce a substantial reorganization of the pore space, as evidenced by the steep initial drop of $C(t)$ from unity.
Panel (b) further shows that changes in the pore space autocorrelation accompany molecular displacements of methane and hydrogen. Therefore, the pore-network cannot be treated as a single fixed graph extracted from one arbitrarily chosen snapshot and then directly matched to a gas trajectory. Instead, the structure must be interpreted statistically: individual pores and throats fluctuate in time, whereas their aggregate distributions provide robust descriptors for trajectory analysis.

The inherent complexity of these systems necessitates trajectory analysis algorithms that account for both molecular motion and structural information about the evolving pore space. This motivates the structure-informed approach developed in the following section, where stationary pore-scale distributions extracted from the dynamic kerogen matrix are used to classify trajectory steps into intra-trap motion and inter-trap transitions.

\section{Trajectory Segmentation and Inference Algorithms}
This section introduces three trajectory labeling classifiers implemented within our open-source library: the conventional distance matrix (DM) algorithm~\cite{lanoiselee2021detecting} used as a baseline, the proposed structure-informed Bayesian (SIB) method, and the hybrid (HYB) approach, which dynamically combines the strengths of both SIB and DM. 
All three methods share a unified common input-output interface. The input consists of a molecular trajectory $\{\mathbf{x}_i\}_{i=1}^{N}$, with $\mathbf{x}_i \in \mathbb{R}^3$, as input.

They process this trajectory differently. DM constructs a pairwise distance matrix from the trajectory points. SIB uses the lengths of consecutive trajectory steps, and HYB combines the outputs of DM and SIB.

The output is a binary trap state sequence $\{y_i\}_{i=1}^{N-1}$, where $y_i = 1$ indicates intra-trap motion and $y_i = 0$ indicates an inter-trap transition.
The following subsections describe the three methods.

\subsection{Geometric Baseline} 
\subsubsection{The Pairwise Distance-Matrix Approach} \label{struct_approach}

\begin{figure*}[t]
\centering
\includegraphics[width=0.90\textwidth]{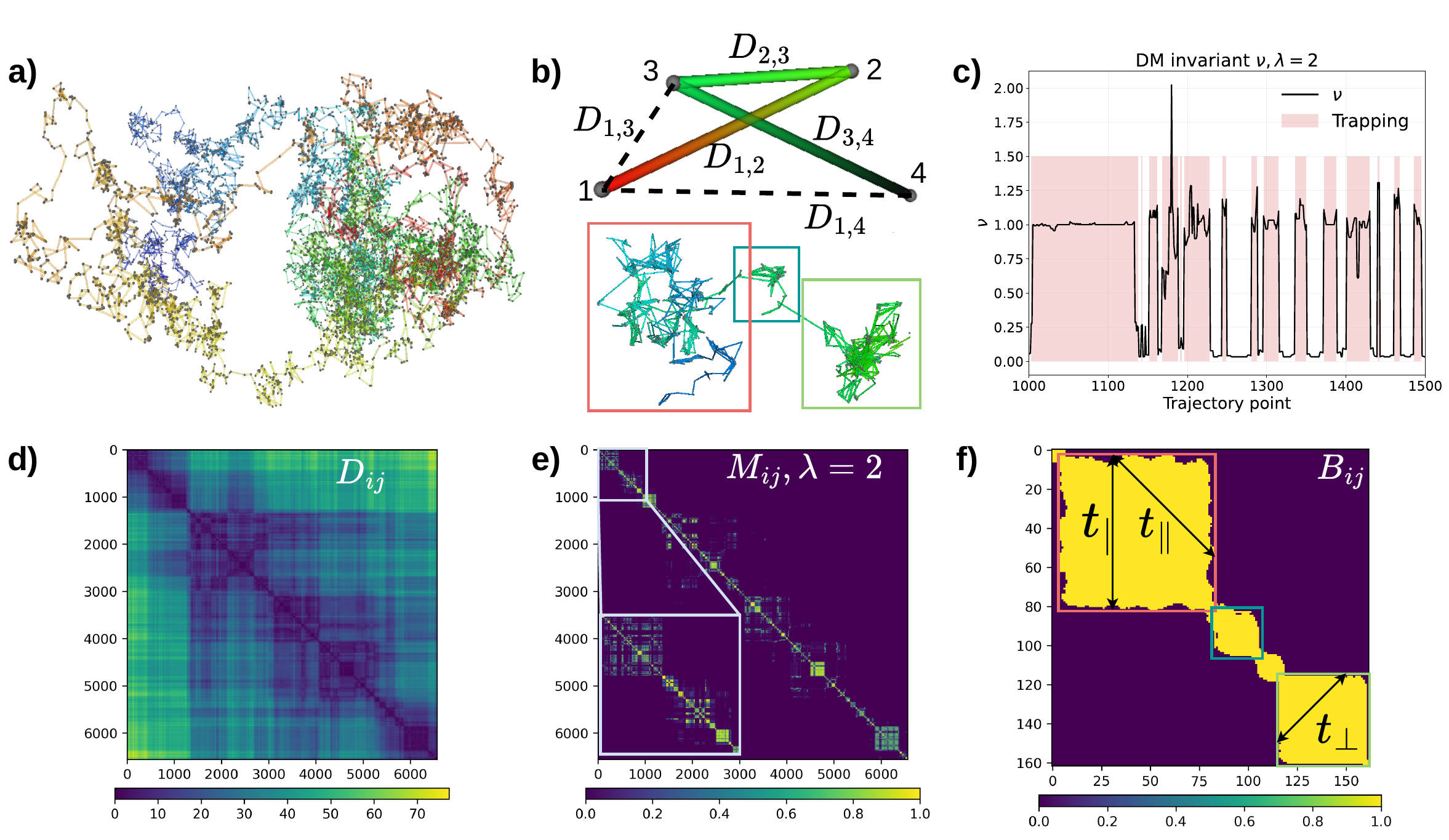}
\caption{ Distance-matrix-based analysis of molecular trajectories and block-invariant descriptors of trapping events. 
(a) Example of a complete trajectory of a single gas molecule. Trajectory color encodes cumulative path length from the starting point.
(b) Schematic construction of the distance matrix $\mathbf{D}$ for a representative trajectory with four points (upper subpanel) and a selected fragment of the full trajectory from panel~(a) with identified trapping events (lower subpanel). Colored frames in the lower subpanel mark trajectory segments that correspond to the diagonal blocks visible in panel~(f). Dashed lines in the upper subpanel indicate auxiliary pairwise distances between nonconsecutive trajectory points.
(c) Distance-matrix block invariant $v(n)$ computed along the trajectory fragment. The curve shows $v(n)$ as a function of trajectory point index for the distance threshold $\lambda$. Shaded regions indicate intervals classified as trapping events, where $v(n)>\nu_c$.
(d) Distance matrix $\mathbf{D}$ for the full trajectory shown in panel~(a). 
(e) Matrix $\mathbf{M}$ computed from $\mathbf{D}$ via a Gaussian kernel with parameter $\lambda=2$. Compact diagonal blocks reflect recurrent visits to the same spatial region and identify potential trapping events.
(f) Binary recurrence matrix $\mathbf{B}$ for the trajectory fragment shown in the lower part of panel~(b), obtained by thresholding the corresponding fragment of $\mathbf{M}$. The diagonal block structure identifies individual trapping events. The geometric quantities $t_{|}(n)$, $t_{\perp}(n)$, and $t_{\parallel}(n)$, used to compute the invariant $v(n)$ shown in panel~(c), are indicated on a representative diagonal block.
}

\label{fig:StructuralAlgorithm}
\end{figure*}

DM~\cite{lanoiselee2021detecting} identifies trapping events as block structures along the diagonal of a pairwise distance matrix.
The method proceeds in four steps:
(i)~Construct the pairwise distance matrix $\mathbf{D}$.
(ii)~Map distances to similarities via a Gaussian kernel $\mathbf{M}$ with length scale $\lambda$, then smooth the result with a uniform $\mu\times\mu$ convolution to obtain $\widetilde{\mathbf{M}}$.
(iii)~Binarize $\widetilde{\mathbf{M}}$ and set $s$ adjacent off-diagonal bands to unity, yielding the binary recurrence matrix $\mathbf{B}$.
(iv)~ Scan the diagonal blocks of $\mathbf{B}$ with a geometric invariant and identify preliminary trapped points using the threshold $v_c$. Candidate trapping runs are then filtered using a critical length obtained from reference free motion through the statistical parameter $p_\mathrm{val}$.

Let a molecular trajectory consist of $N$ points in three-dimensional space, $\{ \mathbf{x}_1, \mathbf{x}_2, \dots , \mathbf{x}_N\}$, where $\mathbf{x}_i \in \mathbb{R}^3$ denotes the position of the molecule at time step $i$ (Fig.~\ref{fig:StructuralAlgorithm}a).
The first step of the algorithm is the construction of the distance matrix $\mathbf{D}$: $D_{ij} = \|\mathbf{x}_i - \mathbf{x}_j\|$, where $\| \cdot \|$ denotes the Euclidean norm (Fig.~\ref{fig:StructuralAlgorithm}b).
The element $D_{ij}$ measures the spatial distance between the molecule's positions at times $i$ and $j$.
A small value indicates that the molecule has returned to (or remained near) the same spatial region.
A compact diagonal block in $\mathbf{D}$ therefore corresponds to a temporal interval during which the molecule is confined to a local region of space.
Off-diagonal blocks arise when the molecule revisits the same trap at distinct time intervals or switches repeatedly between neighboring traps.
The algorithm identifies confinement exclusively through this recurrence pattern, without any knowledge of the physical environment, pore geometry, or characteristic displacement scales of the material.
For a trajectory of length $N$, $\mathbf{D}$ is an $N \times N$ symmetric matrix, whose elements encode pairwise distances between trajectory points (Fig.~\ref{fig:StructuralAlgorithm}d).

In the presence of trapping, segments of the trajectory corresponding to spatial confinement produce compact diagonal blocks in $\mathbf{D}$, reflecting repeated visits to nearby spatial locations. To enhance the contrast between spatially close and distant trajectory points, the distance matrix is transformed into a proximity (similarity) matrix $\mathbf{M}$ using a Gaussian kernel

\begin{equation}
M_{ij} = \exp \left[-\frac{1}{2}
\left(\frac{D_{ij}}{\lambda}\right)^2\right],
\end{equation}

where $\lambda$ is a characteristic length scale controlling the spatial sensitivity of the method.
The parameter $\lambda$ is chosen according to the typical size of traps or characteristic displacements along the trajectory.

To suppress isolated values with high similarity caused by transient thermal revisits (momentary returns to a previously visited position that do not constitute genuine trapping), $\mathbf{M}$ (Fig.~\ref{fig:StructuralAlgorithm}e) is smoothed by a uniform $\mu\times\mu$ box filter with tunable size $\mu$:
\begin{equation}
\widetilde{M}_{ij} = \frac{1}{\mu^2}\sum_{k,l=0}^{\mu-1} M_{i+k,\,j+l}.
\end{equation}
The smoothed matrix $\widetilde{\mathbf{M}}$ eliminates isolated entries with high similarity produced by transient revisits while preserving the extended diagonal blocks that correspond to genuine trapping events.

Next, the proximity matrix is converted into a binary recurrence matrix $\mathbf{B}$ (Fig.~\ref{fig:StructuralAlgorithm}f) using the threshold $p_c$:
\begin{equation}
 B_{ij}=  
\left\{ \begin{aligned} 
  1, \  & if \widetilde{M}_{ij} > p_c \\
  0, \  & otherwise
\end{aligned} \right.
\end{equation}

Because a molecule may briefly exit a trap and immediately enter it again, the detected block can be slightly shifted off the main diagonal. To bridge such gaps, the $s$ diagonals immediately adjacent to the main diagonal are set to unity in $\mathbf{B}$. 
This is a conservative operation that can only merge blocks that are already adjacent and cannot introduce spurious detections in empty matrix regions. The resulting $\mathbf{B}$ exhibits diagonal blocks of ones whose size encodes the duration of each trapping event (Figs.~\ref{fig:StructuralAlgorithm}f).

The geometric structure of the diagonal blocks of $\mathbf{B}$ is characterized by three local measures $t_{|}(n)$, $t_{\perp}(n)$, $t_{\parallel}(n)$ at each diagonal element $n$ (Fig.~\ref{fig:StructuralAlgorithm}f), from which the block-invariant
\begin{equation}
    v(n) = \frac{t_{|}(n)}{t_{\perp}(n) + t_{\parallel}(n) - 1}
\end{equation}
is computed. Here, $v(n)\approx 1$ (Fig.~\ref{fig:StructuralAlgorithm}c) for an ideal trapping block.
For a recurrence matrix without noise the invariant reaches exactly~1.
Real kerogen trajectories exhibit thermal fluctuations that reduce $v(n)$ below this ideal value.
A working threshold of $v_c=0.5$ is therefore used in practice: this conservative value retains genuine trapping segments despite trajectory noise while discarding transient correlations too short to constitute a coherent trapping event.

For a given length scale $\lambda$, the threshold condition defines the preliminary set of trapped points $T(\lambda)=\{n\vert v(n)>v_c\}$.

The parameter $p_\mathrm{val}$ is a statistical threshold parameter, not a run length measured in points. It is applied to a precomputed distribution of block lengths generated by reference Brownian motion (Bm) or fractional Brownian motion (fBm) to obtain a critical run length for each candidate parameter set. 
Contiguous runs in each $T(\lambda_i)$ whose lengths do not exceed this critical value are discarded. The remaining points define the filtered set $T(\lambda_i)$ at each selected scale. The final trapped set is then
\begin{equation}
    T = \bigcup_{i=1}^{n}T(\lambda_i).
\end{equation}
The algorithm thus involves tunable parameters $\{\lambda_1,\ldots,\lambda_n\}$, $v_c$, $p_\mathrm{val}$, $s$, and $\mu$.

The algorithm natively classifies trajectory \emph{points}: $T = \bigcup_{i=1}^{n} T(\lambda_i)$ is the set of trapped point indices, where $T(\lambda_i)$ is the trapped set for scale $\lambda_i$. To produce the labels for trajectory steps required by the common output format, a step from $\mathbf{x}_i$ to $\mathbf{x}_{i+1}$ is labeled as trapped if both its endpoints belong to $T$, yielding the binary trap state sequence $\{y_i^{\mathrm{DM}}\}$ where
\begin{equation}
    y_i^{\mathrm{DM}} = \mathbf{1}[i \in T \;\wedge\; (i+1) \in T].
\end{equation}

\subsubsection{Cyclic Trapping}

\begin{figure}[h]
\centering
\includegraphics[width=0.45\textwidth]{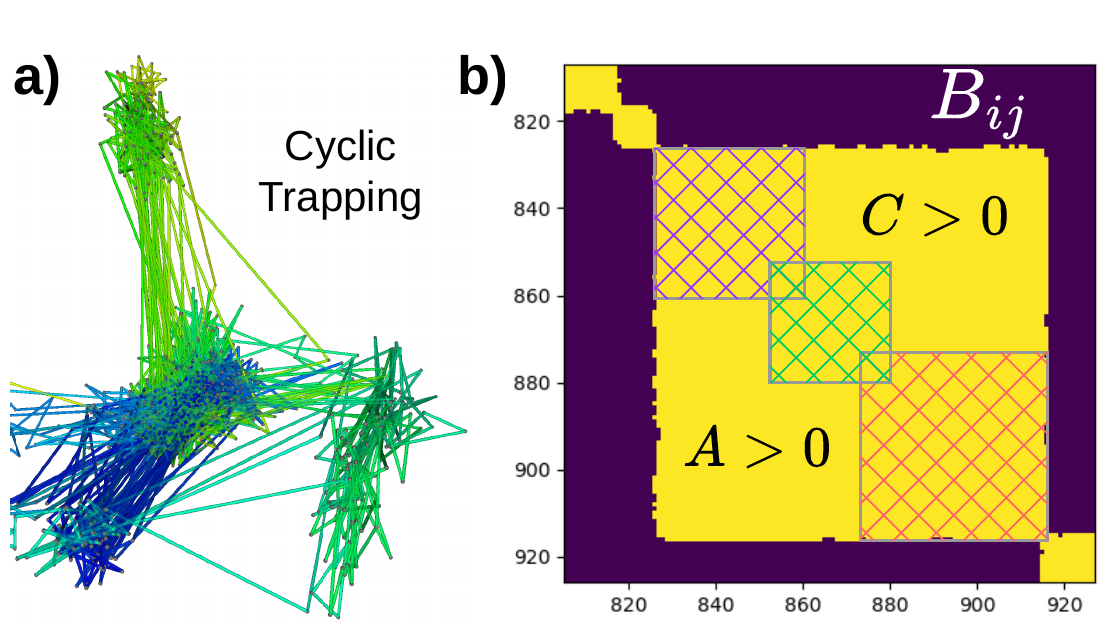}
\caption{ 
Cyclic trapping as a recurrent switching problem in distance-matrix-based trap detection. 
(a) Example of a methane trajectory fragment exhibiting cyclic trapping, where the molecule repeatedly switches between several neighboring traps instead of remaining confined within a single isolated region. The color of each trajectory segment encodes the cumulative path length from the starting point via a sequential colormap that progresses continuously from blue at short accumulated distances to green at larger ones.
(b) Fragment of the transformed distance matrix $\mathbf{B}$ corresponding to the recurrent switching trajectory shown in panel (a). Repeated returns between neighboring traps produce nonzero off-diagonal elements far from the main diagonal ($A>0$, $C>0$), thereby linking otherwise distinct diagonal blocks. As a result, the distance-matrix-based procedure may merge recurrently visited neighboring traps into a single detected region. }
\label{fig:dm_problem}
\end{figure}

Despite its effectiveness in idealized settings, DM has a fundamental limitation: it relies solely on the geometric structure of the trajectory and cannot incorporate physical knowledge of the environment.
A characteristic failure mode in kerogen is \textit{cyclic trapping} (Fig.~\ref{fig:dm_problem}a): the molecule repeatedly switches between several neighboring pores rather than remaining confined within a single pore for a prolonged time.
Each pore pair generates its own diagonal block in $\mathbf{B}$, but repeated returns between pores produce nonzero off-diagonal elements far from the main diagonal (the quantities $A>0$ and $C>0$ in Fig.~\ref{fig:dm_problem}b) that link otherwise distinct blocks.
The algorithm consequently merges all mutually connected blocks into a single detected trapping region, inflating the inferred trapping time.
In kerogen, where pores are coupled through narrow throats and the molecule can traverse the same local pore-network repeatedly, this merging artifact is systematic rather than occasional, rendering DM trap labelling unsuitable for quantitative trapping time analysis.
Detailed derivation and implementation choices of the algorithm are given in the original work~\cite{lanoiselee2021detecting}.
Here we summarize only the components required for comparison with SIB and for mapping its point output to labels for trajectory steps.

\subsection{Structure-informed Bayesian Regularization}

For SIB, the input trajectory is represented by the lengths of consecutive steps, $l_i=\|\mathbf{x}_{i+1}-\mathbf{x}_i\|$, where $i=1,\ldots,N-1$.

Kerogen is a structurally dynamic medium on the timescale of molecular motion.
Thermal fluctuations and local rearrangements continuously modify the connectivity of the pore space, causing regions to transiently open, close, merge, or become disconnected.
As a result, trapping is not a purely geometric property of a static structure, but rather an emergent feature of the coupled dynamics of the matrix and the diffusing molecule.
This observation motivates the use of a structure-informed Bayesian framework, in which structural descriptors of the pore space are incorporated as prior information for classification of trajectory steps.
Because gas molecules diffuse through a connected pore-network and their instantaneous displacements are geometrically constrained by the local pore geometry, pore-network descriptors provide physically motivated external priors: the pore radius distribution characterizes admissible intra-trap displacements, while the throat length distribution characterizes inter-trap transitions.

Although instantaneous pore geometry and connectivity fluctuate, the aggregate pore radius and throat length distributions, $P(r)$ and $P(h)$, remain stationary over the analyzed MD interval and can therefore be used as structural inputs to SIB.

From the pore-network model, the pore radius distribution $P(r)$ and the throat length distribution $P(h)$ are extracted.
From $P(r)$, the intra-trap gas specific step length distribution $\Pi_{\text{gas}}(l)$ is derived and used as the intra-trap likelihood.
$P(h)$ provides the transition length likelihood for inter-trap steps.
The remainder of this section describes these steps and the resulting Bayesian classifier.

\subsubsection{Topology-Preserving Pore-Network Extraction}

\begin{figure*}[t]
\centering
\includegraphics[width=0.9\textwidth]{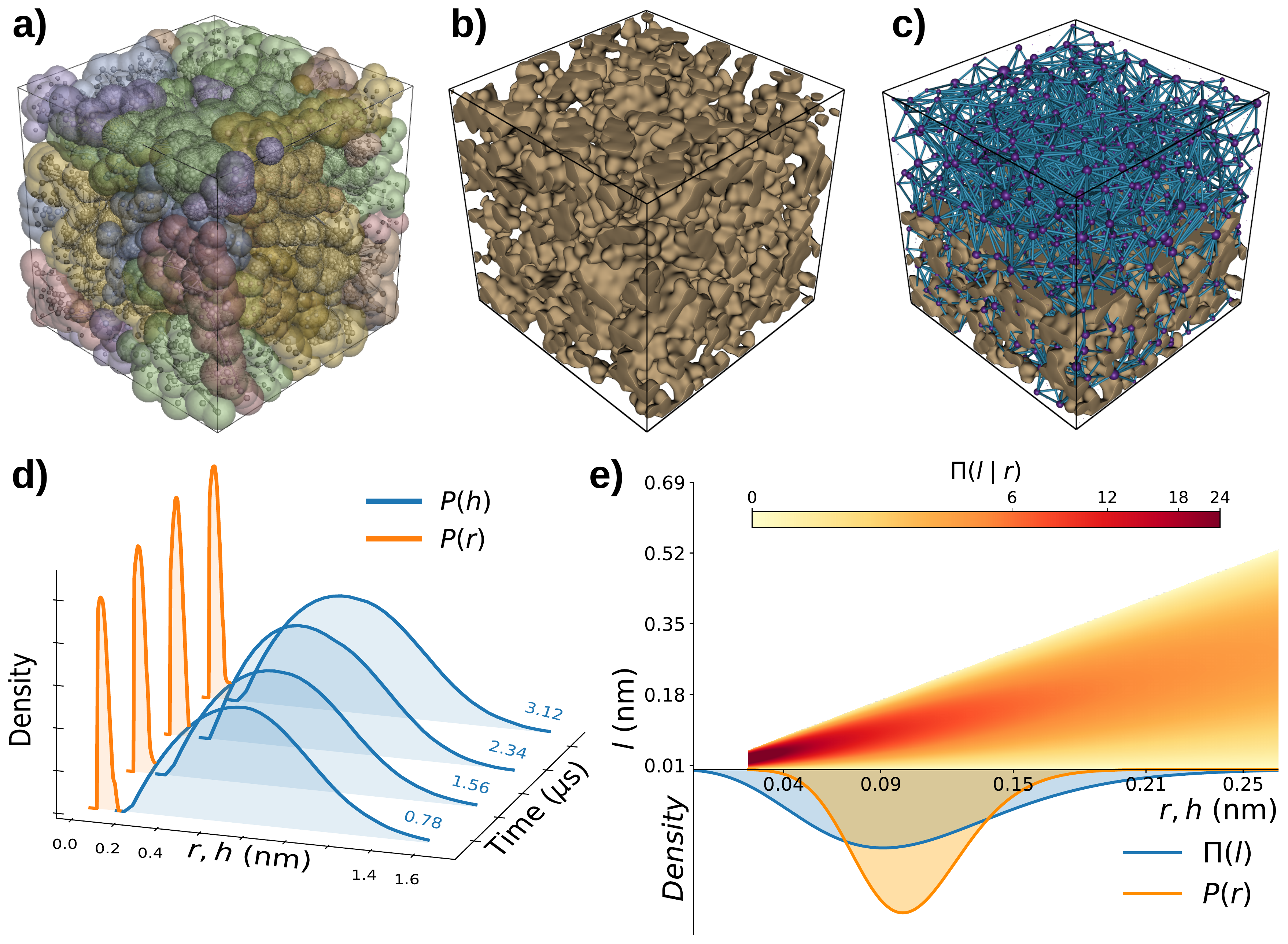}
\caption{ Pore space reconstruction, pore-network extraction, and structural descriptors of the evolving kerogen matrix.
(a) Local subvolume illustrating pore space definition via excluded atomic volumes. Each macromolecule is associated with an individual surface representing the solid phase.
(b) Connected solid matrix obtained after merging individual molecular surfaces. The resulting continuous solid phase defines an interconnected pore space available for gas transport.
(c) The same subvolume as in panels (a) and (b), now with the extracted pore-network model overlaid on the kerogen structure. The lower portion shows the solid phase together with the overlaid pore-network model. The upper portion shows the pore-network model alone, illustrating how the extracted graph spans the accessible void space.
(d) Three-dimensional representation of the pore-network geometric distributions evaluated at different time instants. The curves correspond to the pore radius distribution $P(r)$ and the throat length distribution $P(h)$. Their weak variation over time indicates the statistical stationarity of the pore-scale geometric characteristics.
(e) Construction of the intra-trap step length distribution $\Pi(l)$ from the pore radius statistics. The heat map shows the conditional distribution $\Pi(l\vert r)$ of admissible step lengths inside a pore of radius $r$. The lower panel shows the pore radius distribution $P(r)$ together with the resulting effective step length distribution $\Pi(l)$, obtained by integrating $\Pi(l\vert r)$ over $P(r)$.
}
\label{fig:pnm}
\end{figure*}

To obtain structural descriptors of the kerogen pore space, the atomic coordinates of each MD snapshot are first represented by the exclusion volumes of the kerogen atoms (Fig.~\ref{fig:pnm}a).
Panel~(a) shows how the atomic exclusion volumes associated with each macromolecule contribute to the solid phase.
A uniform voxel grid is placed over the analyzed subvolume, and each voxel is assigned either to the solid kerogen matrix or to the accessible void space.
In practice, a voxel is marked as solid if its center lies within the tabulated atomic radius assigned to at least one kerogen atom,
\begin{equation}
\Phi_{ijk}(t)=
\begin{cases}
1, & \exists a:\|\mathbf{c}_{ijk}-\mathbf{r}_a(t)\|\le R_a,\\
0, & \text{otherwise},
\end{cases}
\end{equation}
where $\mathbf{c}_{ijk}$ is the voxel center position, $\mathbf{r}_a(t)$ is the position of atom $a$ at time $t$, and $R_a$ is the tabulated atomic radius for the chemical element of atom $a$.
Thus, $\Phi_{ijk}(t)=1$ denotes the kerogen phase, whereas $\Phi_{ijk}(t)=0$ denotes the pore space available for gas motion (Fig.~\ref{fig:pnm}b).
Panel~(b) shows the resulting binary image, in which the union of all atomic exclusion volumes forms a connected kerogen matrix that delineates the accessible pore space.
The grid resolution is chosen to resolve narrow pores and throats, and the same voxelization procedure is applied to a sequence of MD snapshots in order to characterize the temporal variability of the pore structure.

The binary image is then reduced to a pore-network model, where pore bodies are represented as graph nodes and throats connecting neighboring pores are represented as graph edges. 
To capture these geometric primitives, we execute a cross-disciplinary technology transfer by adapting the Pore Network Modeling (PNM) approach. 
Historically developed to characterize macroscopic fluid transport and structural heterogeneity in millimeter-to-centimeter scale porous media (such as geological core samples), we scale this paradigm down to the sub-nanometer scale, applying it to highly disordered, dynamic molecular networks. 
Translating PNM into the atomistic regime allows us to reduce complex three-dimensional continuum trajectories into computationally efficient movements on a structural graph. 
However, direct down-scaling introduces severe topological challenges: standard macroscopic extraction algorithms, such as maximal-ball or watershed-based partitionings, frequently suffer from artificial oversegmentation when applied to amorphous materials.
Several classes of pore-network extraction methods based on images are commonly used for this purpose, including maximal-ball or maximal-inscribed-sphere algorithms \cite{dong2009pore}, medial-axis and skeletonization approaches \cite{lindquist2000pore}, and watershed-based partitioning of the distance map \cite{gostick2017versatile}.
These methods differ in how they partition the void space and represent pore connectivity.
A key limitation shared by all of them is that the topology of the extracted network is not guaranteed to match that of the original three-dimensional pore space~\cite{zubov2022pore}.

In this work, we utilize a pore-network extraction procedure constructed from the critical simplices of the signed distance transform via discrete Morse theory combined with persistent homology~\cite{zubov2022pore,kulygin2024porescale}.
In standard geometry-based extractions such as maximal-ball, medial-axis, or watershed partitioning, the network graph frequently suffers from artificial oversegmentation and unphysical coordination numbers because no topological consistency with the original void space is enforced. 
For a trajectory analysis framework, such geometric errors are catastrophic: they introduce spurious throat boundaries and disrupt the true connectivity of the media. 
Discrete Morse theory fundamentally resolves this limitation by mathematically guaranteeing that the Euler characteristic, and consequently the global connectivity structure of the extracted graph, strictly matches that of the original binary pore space. 
This topological fidelity is a critical prerequisite for our Bayesian framework. 
Because the throat-length distribution ($P(h)$) enters the SIB classifier directly as the exact likelihood function for inter-trap transitions, any topological artifacts or broken paths during graph extraction would instantly propagate as non-physical biases into step classification, ultimately distorting the inferred transport mechanisms.

The framework therefore uses both network connectivity and pore morphology rather than global topology alone.
Figure~\ref{fig:pnm}c illustrates the pore-network model extracted from a representative structural snapshot: the network graph is superimposed directly onto the accessible void space of the kerogen subvolume, confirming that the extracted nodes and throats capture the actual pore geometry of the simulation cell.
The method yields such a pore-network model for each analyzed snapshot, from which two structural distributions are extracted: the pore radius distribution $P(r)$ and the throat length distribution $P(h)$ (Fig.~\ref{fig:pnm}d).
Their weak temporal variation indicates that the pore-scale geometric characteristics are statistically stationary over the considered MD interval.
Here, stationarity refers to the aggregate ensemble distributions rather than to the instantaneous geometry and connectivity of individual pores and throats. The fixed periodic cell and the absence of macroscopic polymer swelling provide a stable bulk setting.

\subsubsection{Temporal Stationarity of Geometric Priors}

This remarkable temporal stability is not a mere artifact of a restricted observation window, but a direct consequence of the localized, non-swelling nature of the rigid type-I kerogen matrix at the simulated gas loadings. 
Because the gas concentration falls strictly within the infinite-dilution regime, the dynamic trajectories of the individual gas molecules exert negligible back-action on the macromolecular organic network. 
Consequently, while local, instantaneous pore-throat geometries continuously undergo thermal fluctuations (manifesting as structural 'breathing' at the single-cavity scale) their macroscopic ensemble distributions remain invariant over the entire physically relevant simulation timescale. 
This physical decoupling is rigorously validated by our non-parametric two-sample Kolmogorov–Smirnov (KS) testing (Table~\ref{tab:ks_stationarity} in Appendix~\ref{app:stationarity}), where uniformly high adjusted p-values across all baseline and adjacent time intervals confirm the absence of any systematic structural drift or temporal degradation in the underlying pore networks.
This statistical invariance justifies treating the empirical probability density functions $P(r)$ and $P(h)$ as mathematically robust, time-invariant structural priors that remain completely rigid over the Bayesian iteration steps.

\subsubsection{Derivation of Invariant Intra-Trap Step Length Distributions}

The pore radius distribution $P(r)$ extracted from the pore-network model is stationary and therefore serves as a structural input invariant over time.
Rather than being used directly, $P(r)$ underlies the derivation of $\Pi(l)$, the distribution of intra-trap step lengths, where $l$ denotes the Euclidean distance between successive molecular positions while the molecule remains within a single pore.
Since molecular motion within a trap is geometrically constrained by the pore volume, the relevant quantity is the distribution of admissible displacements inside the pore, not the pore radius itself.

Analytically deriving $\Pi(l)$ from an arbitrary pore size distribution $P(r)$ is intractable in general. 
For a pore of radius $r$, however, the conditional distribution $\Pi(l\vert r)$ of admissible step lengths can be obtained geometrically, and integrating over pore sizes gives $\Pi(l) = \int \Pi(l\vert r) P(r) dr$ illustrated in Fig.~\ref{fig:pnm}e, where the heat map shows $\Pi(l\vert r)$ and the lower panel shows both $P(r)$ and the resulting $\Pi(l)$.

Numerically, $\Pi(l\vert r)$ is obtained by sampling a large number of points uniformly inside a sphere of radius $r$ and computing their pairwise Euclidean distances.
Averaging yields the resulting step length distribution $\Pi(l)$.

The resulting distribution $\Pi_{\text{gas}}(l)$ encapsulates the geometric influence of the pore-scale structure on real molecular displacements during intra-trap motion:

\begin{equation}
    \Pi_{\text{gas}}(l) = \int \Pi(l\vert r) P_{\text{gas}}(r) dr
\end{equation}

Here, gas specific $P_{\text{gas}}(r)$ is the effective pore size distribution scaled specifically for the diffusing gas molecule. 
It is inherited directly from the global, matrix-exclusive geometry $P(r)$ by applying a lower bound cutoff based on the kinetic dimension of the target gas. 
For example, because methane is a relatively large molecule, its real MD trajectory cannot physical enter ultra-confined channels. 
Therefore, when analyzing methane, $P_{\text{gas}}(r)$ ignores these small, inaccessible voids. 
For the smaller and more mobile hydrogen molecule, the cutoff is lower, allowing the framework to explore a wider fraction of the porous network. 
This adaptive approach ensures that the Bayesian priors are strictly consistent with the physical space actually available to each specific gas, while still relying on the time-invariant topological skeleton of the host material.

The resulting $\Pi_{\text{gas}}(l)$ is the structural distribution passed to the Bayesian classification stage.

\subsubsection{Two-Stage Neyman–Pearson and Bayesian Inference}

In this section, we formalize the classification of molecular trajectory steps into two mutually exclusive classes: steps occurring within a trap and transitions between traps.

The empirical structural distributions are not used directly in the Bayesian classifier.
Instead, they are represented by fitted parametric densities: the intra-trap step length distribution $\Pi_{\text{gas}}(l)$ is approximated by $f_T(l)$, whereas the throat length distribution $P(h)$ is approximated by $f_C(l)$.
Because $f_C(l)$ is obtained from $P(h)$, errors in throat identification propagate into the transition likelihood and can bias both classification and the diffusion coefficients inferred from it. 
Errors in pore geometry can likewise affect $P(r)$, $\Pi_{\text{gas}}(l)$, and the intra-trap likelihood $f_T(l)$.

Let $l$ denote the length of an elementary trajectory step. 
Let $H_T$ denote the hypothesis that the step occurs within a trap, and $H_C$ the hypothesis that it corresponds to a transition between traps.

The likelihood of observing a step of length $l$ under each hypothesis is given by

\begin{equation}
p(l\vert H_T) = f_T(l), \qquad p(l\vert H_C) = f_C(l).
\end{equation}

In this framework, the empirical distributions are represented by continuous parametric models: the intra-trap step length distribution $\Pi(l)$ is approximated by a Gamma probability density function (PDF), $f_T(l)$, while the throat-length likelihood, $f_C(l)$, is modeled via a Weibull PDF.

The Gamma distribution naturally captures the bounded, unimodal phase space of intra-pore displacements. 
Crucially, the Weibull distribution is selected due to its mathematical flexibility in capturing the highly asymmetric, heavy-tailed nature of throat lengths inherent to heavily disordered and amorphous macromolecular networks, as verified by our explicit pore-network extraction data. 
While these specific functional forms represent physically motivated modeling choices rather than rigid architectural constraints of the algorithm, they provide an optimal balance between low parameterization overhead and exceptional fidelity to the underlying porous topology.

For a trajectory step of length $l_i$, Bayes' theorem gives
\begin{equation}
P(H_T\vert l_i) =
\frac{f_T(l_i)\,P(H_T)}
     {f_T(l_i)\,P(H_T) + f_C(l_i)\,[1-P(H_T)]}.
\end{equation}
Evaluation of this posterior requires the prior $P(H_T)$, which is the fraction of trapped steps and is initially unknown. The iterative procedure therefore requires an initial approximation $P(H_T)=p_0$. We obtain $p_0$ using the Neyman--Pearson procedure described next.

The initial classification is obtained using the Neyman--Pearson likelihood ratio test~\cite{lehmann2005testing}.
For each step length $l$, the likelihood ratio is defined as

\begin{equation}
\Lambda(l) = \frac{f_C(l)}{f_T(l)}.
\end{equation}

The threshold $\eta$ is chosen to control the probability of labelling a trapped step as a transition,
\begin{equation}
\mathbb{P}(\Lambda(l)\ge\eta\vert H_T)=\varepsilon_\mathrm{NP}.
\end{equation}
This false transition error level is fixed at $\varepsilon_\mathrm{NP}=0.01$. The threshold construction and its numerical implementation are detailed in Appendix~\ref{app:np_criterion}.
The classification rule then reads

\begin{equation}
\Lambda(l_i) < \eta \;\Rightarrow\; H_T,
\qquad
\Lambda(l_i) \ge \eta \;\Rightarrow\; H_C.
\end{equation}

This procedure yields an initial estimate of the fraction of trapped steps,

\begin{equation}
p_0 = \frac{1}{N-1} \sum_{i=1}^{N-1} \mathbf{1}(\Lambda(l_i) < \eta),
\end{equation}

which serves as an informed initial approximation for the Bayesian prior probability $P(H_T)$.

Let $p_k$ denote the estimate of $P(H_T)$ at iteration $k$, initialized with $p_0$ obtained from the Neyman--Pearson step.

For a given $p_k$, posterior probabilities are computed as

\begin{equation}
P_k(H_T\vert l_i) =
\frac{f_T(l_i)\,p_k}
     {f_T(l_i)\,p_k + f_C(l_i)\,(1-p_k)}.
\end{equation}

Under equal misclassification costs for the two hypotheses, the minimum risk decision rule leads to

\begin{equation}
P_k(H_T\vert l_i) > 0.5.
\end{equation}

The prior is then updated as

\begin{equation}
p_{k+1} = \frac{1}{N-1}
\sum_{i=1}^{N-1} \mathbf{1}\!\left(P_k(H_T\vert l_i)>0.5\right).
\end{equation}

Iterations continue until convergence:

\begin{equation}
\left|p_{k+1}-p_k\right|<\delta_\mathrm{conv}.
\end{equation}
We use $0<\delta_\mathrm{conv}\le 1/(N-1)$, which makes convergence equivalent to $p_{k+1} \approx p_k$.
After convergence, let $q_i=P(H_T\vert l_i)$ denote the posterior probability for step $i$, where $i=1,\ldots,N-1$.

This combined Neyman--Pearson and Bayesian framework provides a robust mechanism for trajectory step classification that is consistent with the data and leverages both optimal hypothesis testing and probabilistic refinement.
The output of the algorithm is the binary trap state sequence $\{y_i^{\mathrm{SIB}}\}_{i=1}^{N-1}$, where
\begin{equation}
y_i^{\mathrm{SIB}}=\mathbf{1}\!\left[q_i>0.5\right].
\label{eq:sib_labels}
\end{equation}

In computational practice, the iterative algorithm demonstrates exceptional numerical stability and converges rapidly, consistently requiring only 3 to 5 iterations to reach the predefined convergence threshold $\delta _{conv}$. 
This rapid convergence is mathematically underpinned by our two-stage architecture: the initial Neyman-Pearson likelihood-ratio test provides an analytically optimal initialization point $p_{0}$ that locks the system under strict Type-I error constraints. 
Because $p_{0}$ serves as an exceptionally accurate proxy for the true fraction of trapped steps, the subsequent Bayesian regularization operates in a well-behaved posterior optimization landscape, executing fine-grained probabilistic adjustments rather than large-scale stochastic searches. 
Consequently, the algorithm converges monotonically to a unique, stable fixed point, ensuring that the final trajectory partitioning remains entirely robust to local fluctuations and decoupled from the initialization boundaries.

\begin{figure*}[t]
\centering
\includegraphics[width=0.9\textwidth]
 {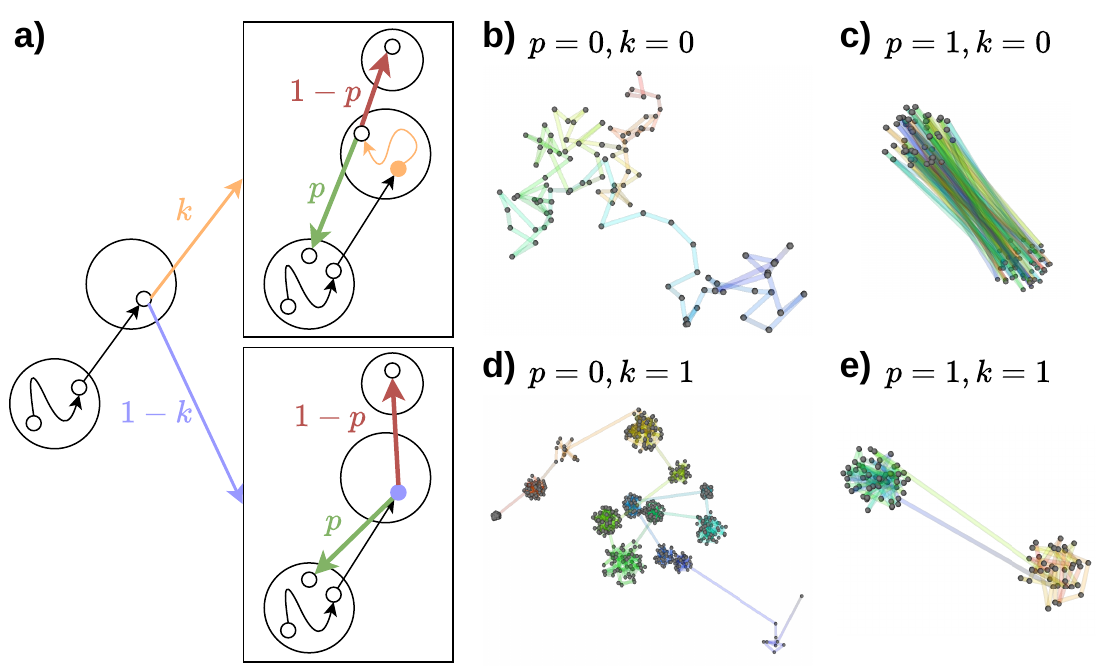}
\caption{
Trajectory simulation model: decision sequence and limiting cases.
(a) Sequential decision diagram for a molecule arriving at a
trap.
Each time the molecule reaches a trap, two decisions are made in sequence.
First, the trapping decision: with probability $k$ (trapping probability) the molecule is captured and performs $s$ intra-trap steps, with $s$ sampled from $P(s)$. 
With probability $1-k$, it bypasses the trap and $s=0$.
Second, the destination decision: with probability $p$ (return probability) the molecule moves to a previously visited adjacent trap. 
With probability $1-p$, it moves to a new trap.
(b–e) Synthetic trajectories at limiting parameter values.
(b) $p = 0.0$, $k = 0.0$: all transitions to new traps, none of the visits to a trap leads to capture.
(c) $p = 1.0$, $k = 0.0$: all returns to previously visited adjacent traps, none of the visits to a trap leads to capture.
(d) $p = 0.0$, $k = 1.0$: all transitions to new traps, every visit to a trap leads to capture.
(e) $p = 1.0$, $k = 1.0$: all returns to previously visited adjacent traps, every visit to a trap leads to capture.
In panels (b–e), trajectory color encodes cumulative path length from the starting point.
}
\label{fig:limit_cases_simulation_model}
\end{figure*}

\subsection{Hybrid Verification for Overlapping Transport Likelihoods}

While the structure-informed Bayesian (SIB) classifier demonstrates exceptional fidelity by operating on stationary physical priors, its localized step-by-step nature inherently encounters a decision bottleneck within the phase-space regions where the intra-trap $\Pi_{\text{gas}}(l)$ and inter-trap $P(h)$ distributions overlappingly exhibit similar length scales. 
In these highly non-ideal, boundary transport regimes the posterior probability $q_{i}$ hovers near the critical decision boundary ($q_i \approx 0.5$), inducing programmatic ambiguity.

To cut through this localized uncertainty, the hybrid (HYB) approach establishes a cross-verification buffer. 
By dynamically combining the strict local probabilistic bounds of SIB with the global recurrence history encoded within the pairwise distance matrix (DM), HYB acts as an automated arbitrator for ambiguous trajectory segments. 
When SIB flags a step as fundamentally uncertain within a tailored ambiguity window $\epsilon_{\text{HYB}}$, the algorithm references the macro-scale recurrence pattern to double-check and correct the local classification, providing a vital physical safety net at a quadratic computational cost.
The value used for the analyses reported in Section~V is $\varepsilon_\mathrm{HYB}=0.3$. 
A step is treated as ambiguous when
\[
\left|q_i-0.5\right|<\varepsilon_\mathrm{HYB}.
\]

The HYB labels are selected for $i=1,\ldots,N-1$ as
\[
y_i^{\mathrm{HYB}} =
\begin{cases}
y_i^{\mathrm{DM}}, & \left|q_i-0.5\right|<\varepsilon_\mathrm{HYB},\\
y_i^{\mathrm{SIB}}, & \left|q_i-0.5\right|\ge\varepsilon_\mathrm{HYB}.
\end{cases}
\]

Although this keeps the cost of HYB quadratic $O(N^2)$, it provides a vital safety net.
A quantitative comparison is provided in Appendix~\ref{app:complexity}.

\section{Stochastic Simulation and Ground-Truth Benchmarking}

A fundamental limitation in analyzing atomistic molecular dynamics (MD) trajectories is the intrinsic absence of objective, step-wise ground truth labels that unambiguously distinguish intra-trap localization from inter-trap transitions. 
In highly disordered amorphous networks, where complex topological features induce severe cyclic trapping artifacts, this lack of baseline labeling prevents a direct, controlled quantification of algorithmic classification errors. 
To overcome this challenge and establish a rigorous validation environment, we introduce a parameterized stochastic trajectory simulation model (TSM). The TSM acts as a mathematically transparent testing ground by generating synthetic molecular trajectories with fully controllable trapping and transition probabilities alongside exact, structurally mapped segment labels. 
By systematically varying the underlying transport parameters, this model provides a robust, physics-informed benchmark to evaluate and contrast the fidelity of the distance-matrix (DM), structure-informed Bayesian (SIB), and hybrid (HYB) classifiers against a known statistical baseline.

The model uses two random decisions to describe molecular motion within and between traps. 
Each time the molecule reaches a trap, two sequential decisions are made (Fig.~\ref{fig:limit_cases_simulation_model}a).
First, the trapping decision: with probability $k$ (trapping probability) the molecule is captured, the number of intra-trap steps $s$ is sampled from $P(s)$, and the molecule performs these $s$ steps. 
With probability $1{-}k$, it bypasses the trap and $s=0$.
Second, the destination decision: with probability $p$ (return probability) the molecule moves to a previously visited adjacent trap. 
With probability $1{-}p$, it moves to a new trap.
These mechanisms are implemented through an evolving trap graph that stores visited traps and their connectivity.

Crucially, the model does not rely on sampling from analytical or averaged probability distributions like $P(r)$ or $P(h)$. 
Instead, it maps the trajectory directly onto the extracted Pore Network Model (PNM) graph of the kerogen cell. 
The simulation engine reads the exact spatial coordinates, individual pore sizes, and localized channel lengths directly from the actual matrix geometry. 
When a molecule moves from one node to another on this graph, its step lengths and confinement boundaries are dictated by the real local dimensions of that specific pore-throat junction. 
This structural grounding ensures that the synthetic trajectories preserve the complete topological complexity and spatial correlations of the genuine amorphous medium. While the spatial parameters of the trajectory are read directly from the exact geometry of the extracted PNM graph, the temporal component follows a different logic. 
Specifically, the distribution $P(s)$, which describes the number of steps performed within a single pore domain, is intentionally kept as a Poisson distribution.
This choice serves as a rigorous, memoryless null model. 
By adopting a standard Markovian decay for intra-pore residence times, we ensure that any observed failures in trajectory partitioning are completely decoupled from temporal heavy-tailed statistics and arise exclusively from spatial and connectivity features, such as cyclic trapping. 
This phenomenological baseline isolates the geometric vulnerability of conventional methods, though \(P(s)\) can be seamlessly updated with empirical non-Markovian or power-law residence time statistics when available.

The trajectory is generated by repeating the following three steps until the prescribed output length $N$ is reached.
\begin{enumerate}
  \item \textbf{Capture decision.} A random number $\xi$ is sampled uniformly from $U(0,1)$. If $\xi < k$, the molecule is captured: the number of intra-trap steps $s$ is sampled from $P(s)$, and the next $s$ trajectory steps are generated inside the current trap sphere of radius $r$. 
  If $\xi \ge k$, the molecule bypasses the trap and no intra-trap steps are generated ($s=0$).
  \item \textbf{Trap selection.} After capture or bypass, a second random number $\xi'$ is sampled uniformly from $U(0,1)$. If $\xi' < p$, a previously visited adjacent trap is selected from the graph. 
  If $\xi' \ge p$, the next destination is designated as a new trap.
  \item \textbf{Transition.} If a previously visited adjacent trap was selected in step~2, a point is sampled uniformly inside that trap and the molecule moves to it. 
  If a new trap is chosen, the particle transitions along an unvisited edge of the current graph node. 
  The displacement $h$ and the radius $r$ of the destination trap are read directly from the specific throat length and pore size attributes stored within that edge and node of the PNM structure. 
  The molecule moves to the exact spatial center of this newly discovered pore room, updating the trajectory history with real geometric parameters.
  In both cases, the displacement to the selected trap is labeled as an inter-trap transition.
\end{enumerate}

Each cycle therefore adds either $s$ intra-trap steps or no intra-trap steps ($s=0$), followed by a transition to the selected trap.
A return reuses an existing graph node, whereas a new destination adds one node and one connecting edge.
Before generation of the recorded trajectory, the same cycle consisting of three steps is used to produce a short prehistory with $p=0.5$.
During this phase, both returns to previously visited adjacent traps and transitions to new traps occur, creating an initial connected trap graph.
The graph and the particle state reached at the end of the prehistory are retained during the subsequent trajectory generation.
The prehistory points are removed only when the final trajectory is returned.

The trapping probability $k$ controls capture at each trap visit: when $k=1$, every visit leads to capture and $s$ is sampled from $P(s)$. 
When $k=0$, every visit is bypassed and $s=0$.
The return probability $p$ controls the next destination: when $p=0$, the molecule always moves to a new trap. 
When $p=1$, it always moves to a previously visited adjacent trap, so no new trap nodes are added.
The qualitative effects of the limiting cases are illustrated in Fig.~\ref{fig:limit_cases_simulation_model}b--e.
Panels (b) and (d) show trajectories containing only transitions to new traps ($p=0$), with bypassing in panel~(b) ($k=0$) and capture in panel~(d) ($k=1$).
Panels (c) and (e) show returns to previously visited adjacent traps ($p=1$), with bypassing in panel~(c) ($k=0$) and capture in panel~(e) ($k=1$).

\section{Results}

\subsection{Algorithmic Performance and Benchmark Validation}

\begin{figure*}[!ht]
\centering
\includegraphics[width=0.75\textwidth]{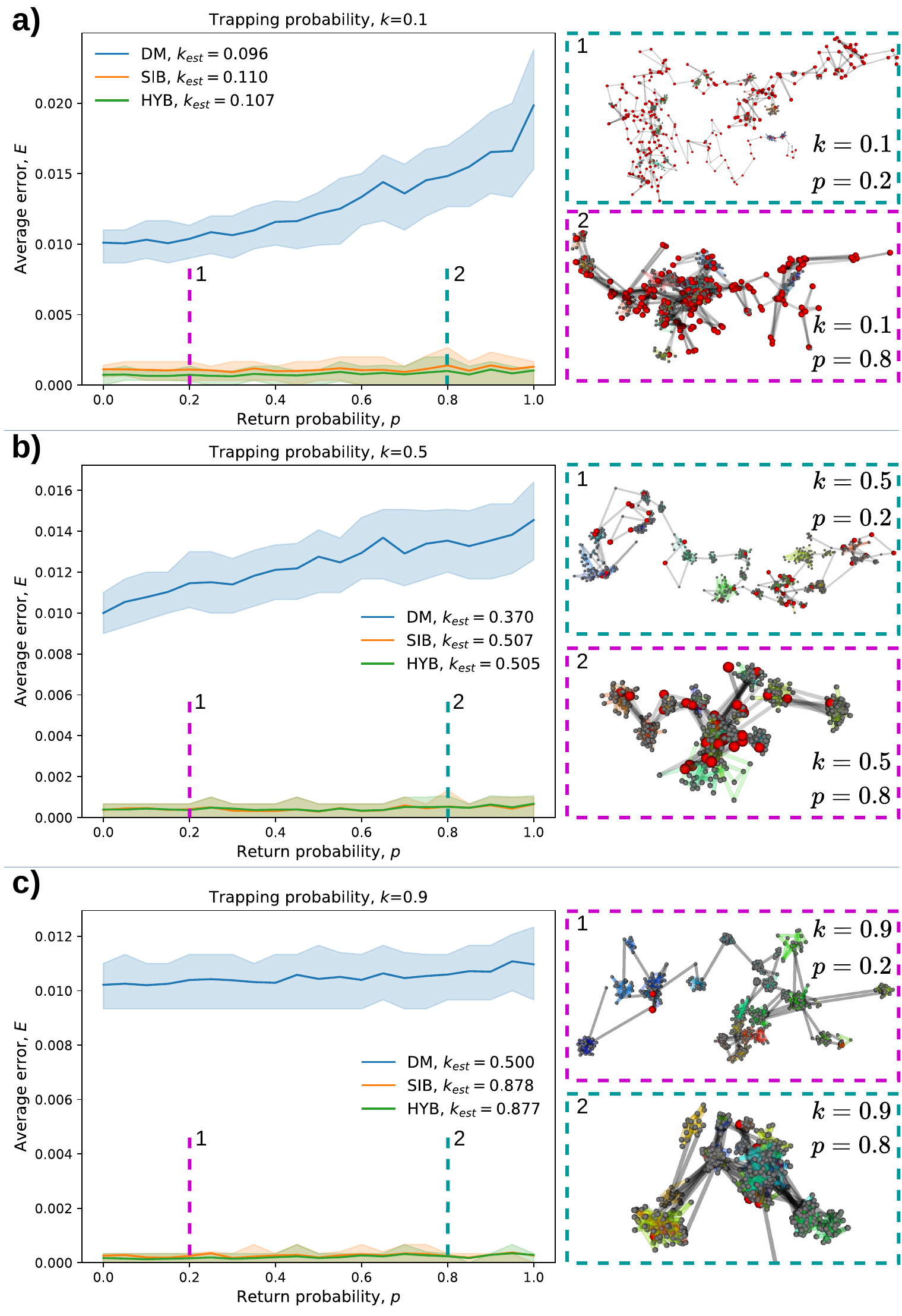}
\caption{
 Evaluation of DM, SIB, and HYB on synthetic trajectories generated by TSM.
Panels (a–c) correspond to different values of the trapping probability $k$: (a) $k=0.1$, (b) $k=0.5$, and (c) $k=0.9$.
 In each panel, the left subpanel shows the average classification error $E$ as a function of the return probability $p$ for DM, SIB, and HYB.
 For each $(p,k)$, the curves show the mean of $E$ over 100 synthetic trajectories of 3000 points. The shaded envelopes span the corresponding 20th to 80th percentile range and include the mean where skewness places it outside that equal-tail interval.
Vertical dashed lines mark selected values of $p$, with corresponding representative trajectory examples shown in the right subpanels.
In the right subpanels, distinct colors identify individual traps (intra-trap motion), gray segments indicate inter-trap transitions, and red spheres mark bypass events with zero trapping time.
}
\label{fig:alg_errors}
\end{figure*}

This subsection evaluates DM, SIB, and HYB on trajectories generated by TSM with known ground truth trap state labels.
The objective is to quantify classification performance across the trapping and return regimes controlled by $k$ and $p$, respectively.
All three classifiers produce a binary trap state sequence $\{y_i^*\}$, where $y_i^* = 1$ indicates intra-trap motion and $y_i^* = 0$ an inter-trap transition.
In this benchmark, $k$ represents the probability of capture at each pore node visit, while $p$ controls the probability of returning to a previously visited adjacent pore room.
Instead of using artificial spatial profiles, the synthetic trajectories were generated by moving the particle directly through the realistic, interconnected Pore Network Model (PNM) graph extracted from the actual kerogen cell. 
This setup ensures that every step length and confinement boundary precisely reproduces the real local distances of the material's channels. 
While the spatial coordinates are strictly governed by the physical graph, the number of steps within a trap was sampled from a standard Poisson distribution $P(s)$ to serve as a memoryless baseline. 
For each combination of $p$ and $k$, 100 independent trajectories of 3000 points were generated. 
The benchmark tested regimes with $k = 0.1$, $0.5$, and $0.9$, while p was varied over the entire range. 
Finally, DM, SIB, and HYB were applied to every trajectory, and each predicted binary trap state sequence $\{y_i^*\}$ was compared directly with its ground truth sequence $\{y_i\}$ using the relative classification error $E$.

The primary metric for algorithm evaluation was the relative error, defined as the fraction of steps misclassified by the algorithm:
\begin{equation}
E = \frac{1}{N-1}\sum_{i=1}^{N-1} \mathbf{1}[y_i^* \neq y_i],
\end{equation}
where $\{y_i\}$ is the ground truth binary trap state sequence and $\{y_i^*\}$ is the predicted binary trap state sequence.
Because the synthetic trajectories provide ground truth labels for transitions between traps and intra-trap motion, $E$ directly measures classification performance.
It is important to note one statistical effect  when looking at the step-wise error $E$ in high-trapping regimes like $k = 0.9$. 
Here, the moving molecule spends almost all of its time vibrating inside the pores (class 1), and only a tiny fraction of steps represents true jumps between pores (class 0). 
Because of this huge imbalance, if an algorithm like DM mistakenly merges two neighbor pores into one, its overall error $E$ will still look deceptively small. 
It simply benefits from guessing the dominant "in-pore" state. However, while the local step error stays low on the plots, the true physics of the transport network is broken, leading to the massive underestimation of macroscopic transition events shown in Table~\ref{tab:k_est_synthetic}. 
This shows why a low step-wise error alone cannot guarantee that the physical mechanism is captured correctly.

In addition to the classification error, synthetic trajectories provide access to the ground truth classification of transitions into trapping events and events without trapping.
This enables a direct estimation of the empirical trapping probability fraction $k_\mathrm{est}$ from the trajectory itself.
\[
k_\mathrm{est} = \frac{N_t}{N_0 + N_t}.
\]

Here, $N_t$ is the number of capture events and $N_0$ is the number of bypass events with zero trapping time.
In the TSM, capture occurs with probability $k$ at each trap visit, so $k_\mathrm{est}$ provides an empirical estimate of the TSM trapping probability $k$.

The classification error $E$ remains the primary measure of label accuracy, while $k_\mathrm{est}$ provides a complementary diagnostic of whether a classifier preserves capture and bypass statistics.
The two TSM parameters play distinct roles.
The parameter $k$ directly sets the capture probability at a trap visit and therefore strongly controls the number of trapping events.
The parameter $p$ determines whether an inter-trap transition returns to a previously visited adjacent trap or enters a new trap, and therefore controls cyclic trapping.
The return probability $p$ cannot be estimated directly from classified MD trajectories.
Identifying the two types of transition requires the identities and connectivity of the visited traps.
This independent ground truth trap graph is unavailable for the MD trajectories, so estimating $p$ would require additional assumptions about the trap structure.

The DM parameters were selected using a brute-force search over the candidate parameter grid to minimize the relative classification error $E$ (Appendix~\ref{app:dm_search}).
SIB requires no analogous parameter search.
HYB uses the corresponding optimized DM parameter set in its structural component and was not optimized separately using the HYB classification error.
The ambiguity threshold was fixed at $\varepsilon_\mathrm{HYB}=0.3$ for the analyses reported here.

\begin{table}[ht]
\centering

\caption{Empirical trapping probability estimates for the synthetic benchmark.
The relative deviation is $\Delta k=100(k_\mathrm{est}-k)/k$.}
\label{tab:k_est_synthetic}
\begin{tabular}{|c|c|c|c|}
\hline
Algorithm & $k$ & $k_\mathrm{est}$ & $\Delta k$ (\%) \\
\hline
DM & $0.1$ & $0.096$ & $-4.0$ \\
DM & $0.5$ & $0.370$ & $-26.0$ \\
DM & $0.9$ & $0.500$ & $-44.4$ \\
\hline
SIB & $0.1$ & $0.110$ & $10.0$ \\
SIB & $0.5$ & $0.507$ & $1.4$ \\
SIB & $0.9$ & $0.878$ & $-2.4$ \\
\hline
HYB & $0.1$ & $0.107$ & $7.0$ \\
HYB & $0.5$ & $0.505$ & $1.0$ \\
HYB & $0.9$ & $0.877$ & $-2.6$ \\
\hline
\end{tabular}
\end{table}

Fig.~\ref{fig:alg_errors} presents the relative error distributions for DM, SIB, and HYB at three trapping regimes: low trapping probability ($k=0.1$, panel a), intermediate trapping probability ($k=0.5$, panel b), and high trapping probability ($k=0.9$, panel c).
The left side of each panel shows error as a function of $p$, whereas the right side shows representative synthetic trajectories at selected $k$ and $p$.
The trajectory insets are marked by dashed vertical lines at $p=0.2$ (inset~1) and $p=0.8$ (inset~2) in each error plot.
The results show that DM produces higher relative errors across most parameter sets.
In contrast, HYB yields error levels close to those of SIB because its primary classification step is based on the same probabilistic criterion.

The synthetic benchmark reveals a critical dual-mode failure in the distance matrix algorithm (DM), exposing both an integral bias and a dynamic trajectory-level vulnerability. 
First, as summarized in Table~\ref{tab:k_est_synthetic}, DM systematically underestimates the aggregate trapping probability $k_{est}$ across all tested regimes. 
Second, and more fundamentally, the point-wise relative classification error $E$ for the DM method exhibits a severe, systematic escalation as a function of the return probability $p$ (Fig.~\ref{fig:alg_errors}). This strong dependence of the trajectory-level error on $p$ provides a direct mathematical fingerprint of the cyclic trapping artifact.

Physically, high values of $p$ correspond to transport regimes where the diffusing molecule frequently and rapidly switches between adjacent, closely spaced pore cavities rather than escaping into the bulk matrix. 
Because DM operates exclusively on the coordinate-blind geometric recurrence of the trajectory itself, these rapid back-and-forth transitions generate dense, overlapping off-diagonal similarity signatures. 
The conventional algorithm cannot distinguish these rapid inter-pore jumps from genuine intra-pore rattling, causing it to erroneously merge distinct localized states into a single, continuous, and artificially prolonged trapping block. 
Consequently, the point-wise error $E$ of DM reaches its absolute peak in these high-$p$ configurations, demonstrating that purely geometric trajectory analysis inherently degrades when subjected to highly cyclic transport topologies. 
In sharp contrast, the structure-informed Bayesian (SIB) and hybrid (HYB) approaches maintain comparatively stable error bands across the entire spectrum of $p$ values. 
While maintaining an exceptionally low step-wise misclassification error ($E < 0.2\%$), SIB restricts the aggregate relative deviation in the inferred trapping probability ($\delta k$) to a narrow window between $-2.4\%$ and $10.0\%$, effectively decoupling macro-transport statistics from geometric recurrence artifacts.

The computational scaling of DM, SIB, and HYB is analyzed in Appendix~\ref{app:complexity}, where the SIB algorithm is shown to exhibit approximately linear scaling and therefore remains the most computationally efficient choice for large-scale trajectory datasets.
These synthetic results establish the accuracy limits of the classifiers before their application to MD trajectories.

\subsection{Decoding Gas Transport Micro-Mechanics in Real Molecular Trajectories}

\begin{figure*}
\centering
\includegraphics[width=0.98\textwidth]{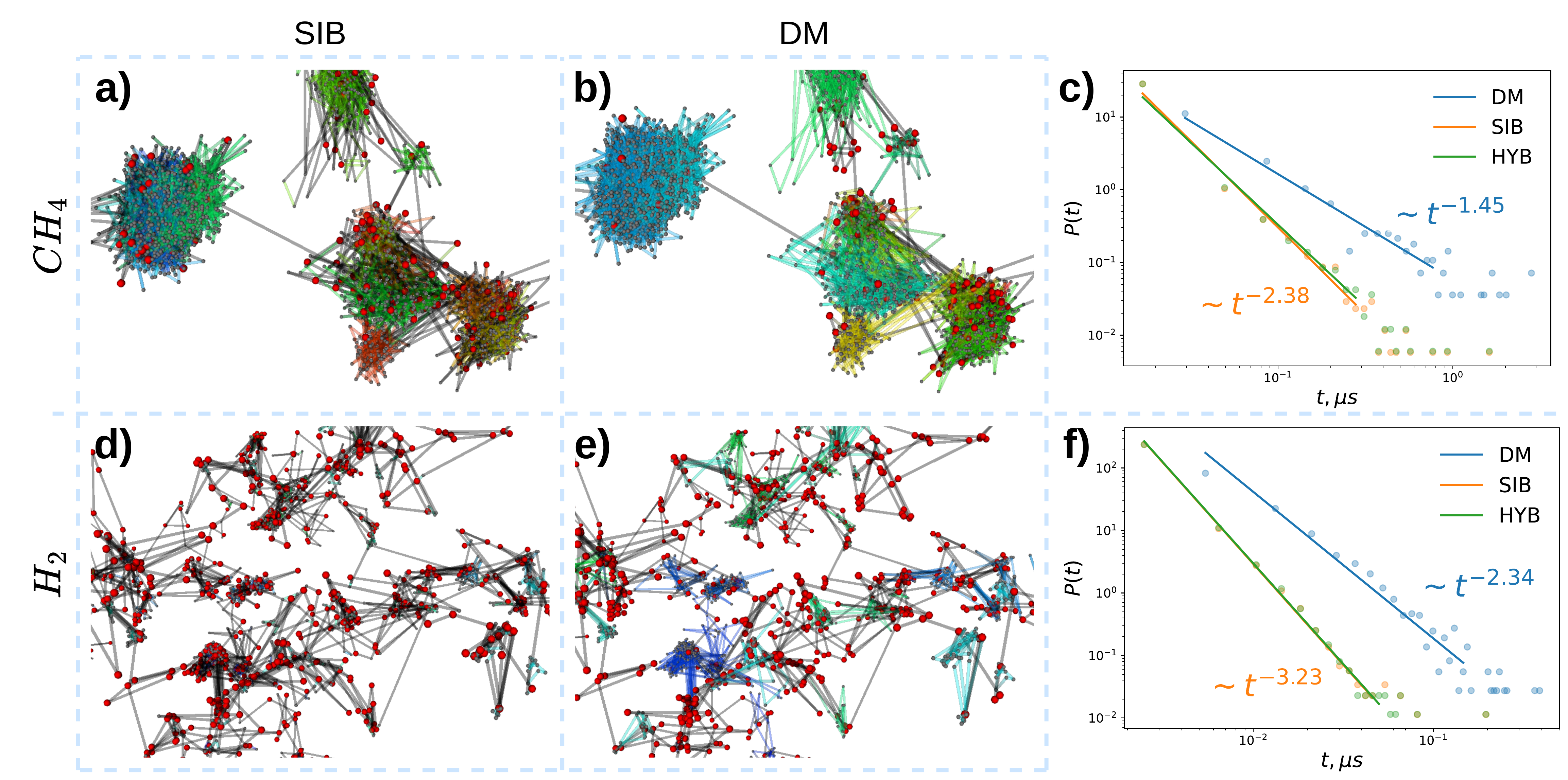}
\caption{
Analysis of trapping time statistics and identification of anomalous diffusion regimes.
Top row (a--c): methane ($\mathrm{CH}_4$). Bottom row (d--f): hydrogen ($\mathrm{H}_2$).
(a,d) MD trajectories classified using SIB.
(b,e) MD trajectories classified using DM.
In panels (a,b,d,e), gray segments indicate inter-trap transitions, colored segments indicate intra-trap motion, and red spheres mark trap visits that did not result in capture.
(c,f) Empirical trapping time distributions $P(t)$ obtained by applying DM, SIB, and HYB to all 20 molecular trajectories, shown on log--log axes.
The scatter points correspond to the distributions obtained using the three methods, while the black lines show the corresponding power-law fits used to estimate the tail exponent $\mu$.
}

\label{fig:time_trapping}
\end{figure*}

This subsection applies all three classifiers to the MD trajectories of methane ($\mathrm{CH}_4$) and hydrogen ($\mathrm{H}_2$) in kerogen.
Since MD provides no ground truth labels, the comparison is qualitative: representative classified trajectories (Figs.~\ref{fig:time_trapping}a,b,d,e) illustrate systematic differences between the classifiers, and the empirical $P(t)$ distributions (Figs.~\ref{fig:time_trapping}c,f) show how classifier choice affects the inferred trapping time statistics.
Based on the synthetic benchmark of Section~V.A, SIB is used as the primary analysis method.
DM is retained as a reference baseline to demonstrate how the conventional approach can merge neighboring traps and distort the inferred trapping time statistics.

For both gases, the primary comparison metric is the empirical trapping time distribution $P(t)$, as it directly characterizes how long molecules remain confined.
The fitted tail exponent $\mu$ quantifies the decay of the density over the selected finite interval.
Smaller values of $\mu$ correspond to a higher relative frequency of long trapping events within that interval.
Therefore, differences in $P(t)$ and $\mu$ between classifiers indicate systematic differences in trapping time detection and in the resulting trapping statistics.

The resulting trapping time distributions $P(t)$ highlight differences in the capture dynamics of methane and hydrogen, reflecting their different molecular sizes and interaction profiles with kerogen.
Fig.~\ref{fig:time_trapping}a, b, d, e shows representative classified MD trajectories.
Panels (a) and (d) correspond to SIB applied to methane and hydrogen, respectively, while panels (b) and (e) show the corresponding DM classifications.
These panels illustrate how DM can merge neighboring traps, whereas SIB assigns shorter trap segments that better follow the local pore-scale organization of the trajectory.
All three classifiers were used to calculate $P(t)$ and the scalar statistics reported below. 
Representative classified trajectories are shown only for DM and SIB. 
HYB is not shown in separate panels since its classifications are nearly identical to those of SIB.

Fig.~\ref{fig:time_trapping}c and Fig.~\ref{fig:time_trapping}f present the trapping time distributions $P(t)$ on log--log axes for methane and hydrogen, respectively, showing results for DM, SIB, and HYB.
Panel~c corresponds to methane ($\mathrm{CH}_4$): the distribution decays more slowly and has a smaller $\mu$, reflecting a greater relative contribution of long trapping events.
Panel~f corresponds to hydrogen ($\mathrm{H}_2$): the faster decay and larger $\mu$ are consistent with a smaller relative contribution of long trapping events.

This subsection summarizes the scalar statistics extracted from $P(t)$: the power-law exponent $\mu$, the number of trapping transitions $N_t$, and the empirical trapping probability fraction $k_\mathrm{est}$.
Comparing these metrics across algorithms and gases reveals systematic differences in trapping behavior and exposes biases specific to DM.

Table~\ref{table:table_time_trapping} reports $\mu$, $N_t$, $N_0$, and $k_\mathrm{est}$ for all three classifiers applied to MD trajectories of methane and hydrogen.
Here, $\mu$ is the fitted trapping time tail exponent, $N_t$ is the number of classified capture events, $N_0$ is the number of classified bypass events, and $k_\mathrm{est}$ is the estimated capture probability.
To improve statistical reliability, all reported quantities are averaged over an ensemble of 20 independent molecular trajectories for each gas.
For each trajectory $j$, the capture fraction is first calculated as
$k_\mathrm{est}^{(j)}=N_t^{(j)}/(N_0^{(j)}+N_t^{(j)})$; the table reports the ensemble mean $\langle k_\mathrm{est}^{(j)}\rangle$.
The reported $N_t$ and $N_0$ are averaged separately and are therefore not restricted to integers.
Consequently, the ensemble-averaged capture fraction is not, in general, equal to the ratio of the separately averaged event counts,
$\langle k_\mathrm{est}^{(j)}\rangle \neq \langle N_t^{(j)}\rangle/(\langle N_0^{(j)}\rangle+\langle N_t^{(j)}\rangle)$.
Applied to MD trajectories, this trajectory-wise capture fraction provides an empirical estimate of the TSM trapping probability $k$ (cf.\ Section~V.A), with $N_t^{(j)}$ and $N_0^{(j)}$ extracted from algorithmic classification rather than from ground truth labels.

The data reveal significant differences in the performance of the algorithms.
DM shows limited sensitivity to the parameter $k_{est}$, producing almost identical values for methane and hydrogen.
This insensitivity is at odds with the visual analysis of trajectories, which clearly indicates distinct trapping behaviors for the two gases.
Methane demonstrates more prolonged trapping times, while hydrogen exhibits faster transitions and less frequent trapping.
In contrast, the SIB and HYB classifiers yield $k_\mathrm{est}$ values that differ clearly between the two gases, consistent with the qualitative trajectory analysis.

Crucially, the numerical robustness of the decoded power-law tail statistics is fundamentally dictated by the observation window and the density of recorded data. 
As emphasized in statistical studies of empirical power laws, tail fluctuations and the finite range over which scaling holds can strongly bias exponent estimates when data are sparse~\cite{clauset2009power}.
Long-duration trapping events represent classic statistical rare events; because they lie in the extreme far tail of the probability density function, capturing a statistically sufficient ensemble of these extended periods is impossible within localized, short-time windows. 
This sampling bottleneck provides the core motivation for deploying high-performance computational clusters to generate massive, multi-microsecond atomistic trajectories. 
Resolving the true, un-biased transport exponents requires thousands of continuous transitions to ensure that these macroscopic rare events are fully represented rather than lost to finite-size sampling noise. 
Consequently, generating long molecular dynamics ensembles is not a mere computational luxury, but a strict physical necessity to safely overcome tail fluctuations, navigate finite-range scaling limits, and extract stable, mathematically rigorous estimates of the anomalous trapping time exponent.

The metrics summarized in Table~\ref{table:table_time_trapping} characterize the trapping dynamics of different gases and provide a basis for calibrating the $k_{est}$ parameter in synthetic trajectory simulations.
This calibration enables the generation of synthetic trajectories that reproduce the main statistical features of the MD data, providing a controlled benchmark for further studies of molecular transport and trapping mechanisms.

\begin{table}
\centering
\caption{
Trapping time analysis results for methane ($\mathrm{CH}_4$) and hydrogen ($\mathrm{H}_2$) using the DM, SIB, and HYB classifiers.
The positive tail exponent $\mu$ is defined by $P(t) \propto t^{-\mu}$.
}
\label{table:table_time_trapping}
\resizebox{\columnwidth}{!}{%
\begin{tabular}{ |c|c|c|c|c|c| }
\hline
 &  & $\mu$ & $N_t$ & $N_0$ & $k_{est}$ \\
\hline
\multirow{2}{6em}{DM} & $CH_4$ & 1.452 & 24.7 & 40.9  & 0.400 \\
                         & $H_2$  & 2.345 & 233.5 & 669.8 & 0.260 \\
\hline
\multirow{2}{6em}{SIB} & $CH_4$ & 2.381 & 265.0 & 152.9 & 0.671 \\
& $H_2$ & 3.229 & 1118.2 & 2018.4 & 0.358 \\

\hline
\multirow{2}{6em}{HYB} & $CH_4$ & 2.247 & 254.5 & 145.4 & 0.672 \\
& $H_2$ & 3.221 & 1110.3 & 1982.0 & 0.360 \\
\hline
\end{tabular}
}
\end{table}

$N_t$ and $N_0$ count classified trap encounters rather than trajectory points or fractions of the total simulation time. Each contiguous trapping interval contributes one capture event to $N_t$, while its points determine a single duration $\tau$. A bypass contributes one event with zero trapping time to $N_0$. Therefore, $N_t+N_0$ is not expected to equal the number of trajectory steps, and the remaining inter-trap transition steps do not form a third trap encounter category.
For methane, SIB and HYB give similar mean event counts. SIB gives $N_t=265.0$ and $N_0=152.9$, while HYB gives $N_t=254.5$ and $N_0=145.4$. DM gives $N_t=24.7$ and $N_0=40.9$. For hydrogen, SIB gives $N_t=1118.2$ and $N_0=2018.4$, while HYB gives $N_t=1110.3$ and $N_0=1982.0$. DM gives $N_t=233.5$ and $N_0=669.8$. All three classifiers identify substantially more trap encounters for hydrogen than for methane.

\begin{figure}[ht]
\centering
\includegraphics[width=0.37\textwidth]
 {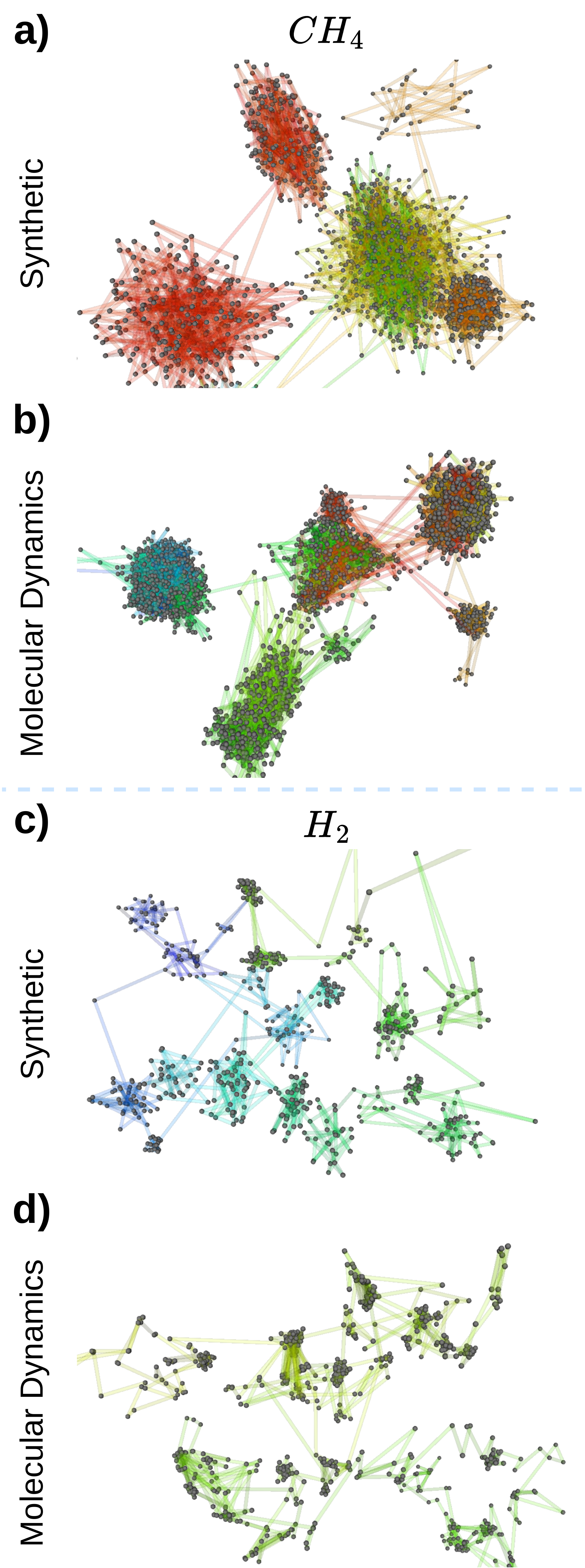}% Here is how to import EPS art
\caption{
Comparison of synthetic trajectories generated by the proposed model with molecular dynamics simulations.
(a) Synthetic trajectory obtained using the proposed algorithm with parameters $p = 0.8$ and $k = 0.67$.
(b) Molecular dynamics trajectory of a methane molecule ($CH_4$), shown for comparison with panel (a).
(c) Synthetic trajectory obtained with parameters $p = 0.2$ and $k = 0.36$.
(d) Molecular dynamics trajectory of a hydrogen molecule ($H_2$), shown for comparison with panel (c).
In all panels, trajectory color encodes cumulative path length from the starting point.
}
\label{fig:sim_vs_real}
\end{figure}

To visually ground the quantitative metrics reported in Table~\ref{table:table_time_trapping} and demonstrate that our stochastic framework captures the fundamental physics of unconventional gas transport, we perform a direct morphological comparison between the MD trajectories and the TSM outputs.
The simulation parameters for the TSM were strictly calibrated against the empirical trapping fractions extracted by the SIB classifier, yielding a high-capture regime ($k=0.67$) for methane and an intermediate-capture, more exploratory regime ($k=0.36$) for hydrogen.
As illustrated in Figure~\ref{fig:sim_vs_real}, the synthetic trajectories generated under these structurally-informed constraints show striking qualitative agreement with the explicit molecular dynamics runs. 
For methane (Figs.~\ref{fig:sim_vs_real}a,b), both the TSM and MD paths manifest as tightly bound, highly localized spatial clusters, where the molecule spends prolonged intervals undergoing intra-pore rattling before executing rare, rapid transitions. 
Conversely, for the smaller and less interacting hydrogen molecule (Figs.~\ref{fig:sim_vs_real}c,d), both frameworks consistently reproduce more open, delocalized trajectories characterized by lower-density clusters and a significantly higher frequency of inter-pore jumps.

The return probability $p$ cannot be estimated directly from the classified MD trajectories because the identity and connectivity of the visited traps are unknown.
It was therefore selected by comparing the spatial distribution of trajectory points, the size and density of localized clusters, the number of successive points within these clusters, and the visually apparent frequency of transitions between clusters.
For methane, the TSM trajectory with $k=0.67$ and $p=0.8$ in panel~a and the MD trajectory in panel~b both contain compact, densely populated clusters.
Many successive trajectory points remain within these clusters, indicating prolonged localized motion, and the apparent numbers of transitions between clusters are comparable.
For hydrogen, the TSM trajectory with $k=0.36$ and $p=0.2$ in panel~c and the MD trajectory in panel~d contain smaller and less densely populated clusters.
They also contain fewer successive points within individual clusters, consistent with shorter residence, while the apparent numbers of transitions between clusters remain comparable.

This comparison is necessarily qualitative.
Without independent trap labels for the MD trajectories, the spatial path alone does not determine unambiguously whether a step remains within a trap or connects two different traps.
Within this limitation, the selected TSM regimes reproduce the contrast between the more persistent localized motion of methane and the more exploratory motion of hydrogen.

Ultimately, this morphological convergence confirms that by utilizing stationary pore-network descriptors as Bayesian invariants, the developed framework does not merely classify steps post hoc, but successfully decodes the genuine, coupled space-time dynamics of anomalous diffusion within complex amorphous topologies.

\section{Discussion}
\subsubsection{The Paradox of Structural Breathing: Stationary Invariants in Dynamic Media}

A central conceptual paradox resolved by our framework lies in the coexistence of highly dynamic, short-term atomic fluctuations, described as structural $\textit{breathing}$, with the strict, long-term statistical stationarity of the aggregate pore-space distributions. 
Related studies of cyclically evolving porous materials have shown that combined morphological and topological descriptors can reveal nonmonotonic structural transformations that scalar measures alone may miss~\cite{tolstygin2026soil}.
Under standard operating conditions, the amorphous macromolecular network of type-I kerogen is far from rigid; thermal motions continuously open, close, and distort local transport pathways on picosecond timescales, creating a highly transient confinement environment for diffusing gas molecules. 
If the polymeric host material were completely static, partitioning single-molecule trajectories into intra-trap localization and inter-trap transitions would reduce to a trivial geometric lookup problem. 
One could simply extract the pore-network graph once from a single structural snapshot and project the trajectory points directly onto the fixed spatial domains of pores and throats.

However, because the amorphous matrix is dynamically fluctuating, such a naive static projection systematically fails. 
At the same time, executing a full, topology-preserving graph extraction via discrete Morse theory at every individual molecular dynamics step is computationally impossible, as the multi-scale geometric voxelization of shifting atomic coordinates would induce an intractable numerical overhead.

Our SIB framework cuts through this computational deadlock by shift-ing the analytical paradigm from localized geometric tracking to global probabilistic regularization. 
While an individual pore cavity dynamically alters its instantaneous shape, the ensemble-averaged distributions of pore radii, $P(r)$, and throat lengths, $P(h)$, remain strictly time-invariant macroscopic properties of the material, as rigorously demonstrated by our Kolmogorov–Smirnov benchmarks. 
By encoding these stationary aggregate distributions directly into the Bayesian step-classifier as invariant priors and likelihood functions, the SIB framework effectively untangles the coupled space-time complexity of transport in soft matter. 
It accommodates the local physical reality of structural breathing without needing to track it explicitly at every femtosecond step, offering an elegant, high-fidelity solution that bridges transient atomistic fluctuations with long-time continuum transport statistics.

Ultimately, this approach transforms the entire treatment of anomalous diffusion in porous media: instead of modeling transport as an abstract, energy-driven CTRW in an open bulk space, our framework ground the physics within a topologically-conditioned random walk on a discrete graph of the material's void network, successfully reconciling local structural breathing with global transport stationarity.

\subsubsection{Methodological Impact and Computational Scalability in the HPC Era}

Beyond its physical fidelity, the developed structure-informed Bayesian framework provides a critical methodological breakthrough that addresses a long-standing computational bottleneck in single-molecule trajectory analysis. 
Traditional geometric segmentation algorithms, most notably the conventional distance-matrix (DM) approach, suffer from a fundamental quadratic complexity scaling, $O(N^2)$, in trajectory length. 
In the current era of high-performance computing (HPC), where state-of-the-art atomistic simulations effortlessly generate massive, multi-microsecond to millisecond single-particle tracking datasets, this quadratic overhead renders the automated processing of large ensembles completely intractable. 
Importantly, because long-lived trapping states behave as statistical rare events in the extreme tail of the PDF, short-time observation windows cannot resolve them without severe finite-size sampling noise. 
This fundamental limitation underscores why utilizing high-performance computing resources to accumulate extensive, multi-microsecond trajectories is not a computational luxury, but a strict physical necessity. 
Thousands of continuous transitions are mandatory to safely overcome tail fluctuations, navigate finite-range scaling limits, and extract mathematically rigorous estimates of the anomalous transport exponents.

A trajectory of length $N$ optimized to resolve heavy-tailed, long-time transport statistics effectively causes geometric matrix methods to saturate available memory and CPU time, severely restricting their utility to short, idealized window sizes.
By reformulating trajectory partitioning as a sequential, memoryless step-classification problem conditioned on spatial likelihood invariants, our SIB framework establishes a new linear-scaling $O(N)$ paradigm. 
The practical implications of this transformation are profound: the SIB framework is uniquely equipped to act as a high-throughput, automated pipeline capable of processing massive simulation datasets from advanced supercomputing architectures or high-performance MD platforms like OpenMM without a prohibitive computational footprint. 
While extracting the structural descriptors requires an initial preprocessing of the pore topology, this step behaves as a one-time investment that scales independently of the ensemble size. 
Consequently, the linear computational scaling achieved here is not merely a quantitative improvement, but a qualitative leap forward. 
It bridges the microscopic-to-macroscopic divide, enabling computational physicists to routinely deploy high-fidelity statistical regularizations across macroscopic timescales while maintaining strict atomistic detail.

\subsubsection{The Intrinsic Blindness of Coordinate-Only Tracking to Cyclic Trapping}

Our benchmark findings reveal a fundamental flaw in purely geometric trajectory segmentation: the point-wise classification error is highly non-robust to the topological return probability of the network. 
The catastrophic increase in the conventional algorithm's error as the return probability approaches unity proves that coordinate-only recurrence signatures are inherently unable to separate spatial localization from cyclic transitions. 
In heavily disordered amorphous media—where rapid switching between adjacent pores is a dominant transport mechanism—conventional geometric frameworks systematically break down, misinterpreting fast inter-pore jumps as prolonged confinement. 
This systematic trajectory-level degradation implies that previous literature utilizing coordinate-blind tracking has likely suffered from an overlooked bias, severely overestimating molecular residence times in highly coupled nanoporous systems.
The vital importance of this failure mode becomes starkly apparent when analyzing the real molecular dynamics of gas transport. 
Our empirical calibration reveals a profound physical divergence between the two gases: while the highly mobile, weakly interacting hydrogen molecule exhibits a low return probability ($p = 0.2$), the larger and more strongly bound methane molecule is heavily governed by a high-recurrent regime ($p = 0.8$). 
Due to higher energetic barriers and geometric confinement, methane frequently executes rapid, cyclic returns to previously visited pore cavities.
Consequently, it is precisely for methane—the primary constituent of natural gas—that traditional geometric frameworks suffer their most severe breakdown. 
By conflating these rapid cyclic returns with continuous, uninterrupted trapping, conventional algorithms artificially inflate the inferred trapping intervals, distorting the true power-law transport exponents. 
Our structure-informed framework effectively neutralizes this systematic bias. 
By conditioning step-probabilities on time-invariant structural priors, it maintains a flat, near-zero error profile regardless of the return frequency, successfully unmasking the true microscopic mechanisms of anomalous transport. 
This transformation proves that physical knowledge of the environment is mandatory to accurately decode molecular transport within dynamically fluctuating nanostructures.

\subsubsection{Bridging the Micro-to-Macro Divide: From Atomistic Trajectories to Continuum Fluxes}

The ultimate value of unmasking hidden trapping states lies in our ability to build a predictive, mathematically rigorous bridge from microscopic single-molecule trajectories to macroscopic engineering scales. 
In practical applications such as optimizing gas recovery from unconventional shale reservoirs or designing advanced polymeric membranes for carbon capture and hydrogen separation, direct atomistic simulations of macro-scale devices are fundamentally blocked by computational complexity. 
Engineers inevitably rely on continuum transport frameworks, such as the continuous-time random walk or fractional diffusion equations, which abstract millions of underlying molecular encounters into a handful of effective macroscopic coefficients.

However, these macro-scale models are highly sensitive to their microscopic inputs, specifically the exact statistics of trapping residence times. 
Prior to this work, the systematic tendency of geometric algorithms to conflate cyclic returns with continuous trapping meant that the input parameters fed into continuum models were inherently distorted, overestimating molecular confinement and leading to deeply flawed predictions of gas fluxes.

By successfully decoupling true intra-pore localization from topological recurrence, our structure-informed Bayesian framework provides the clean, unbiased statistical baseline required to parameterize macroscopic models with strict physical fidelity. 
Related multiscale pore network formulations couple nanoconfined phase behavior to larger scale transport through classical density functional theory~\cite{nesterova2025bridging}.
For instance, the stark mechanical contrast we decoded between methane and hydrogen trapping dynamics can be directly translated into macro-scale transport equations to accurately predict the selective permeability of gas separation membranes. 
Consequently, the framework presented here is not merely an analytical tool for trajectory processing, but a generative methodology. 
It provides computational physicists and materials engineers with a transparent, high-throughput pipeline to convert noisy, microsecond-scale atomistic simulations into reliable, long-time continuum predictions, directly advancing our capacity to design next-generation materials for unconventional energy resources and the emerging hydrogen economy.

\section{Conclusion}
This work reframes molecular trajectory analysis in breathing nanopores as a problem of inference from persistent structural statistics rather than reconstruction of transient local geometry.
The key physical finding is that dynamic disorder does not preclude a stable structural description.
Although individual pores and throats continuously change, two-sample Kolmogorov--Smirnov tests with Holm--Bonferroni correction detected no significant temporal drift in the aggregate pore-radius and throat-length distributions, $P(r)$ and $P(h)$, over the simulated interval.
Local structural breathing and stationary ensemble geometry can therefore coexist, allowing the latter to act as a physical regularizer for transport through the evolving matrix.

The structure-informed Bayesian (SIB) framework operationalizes this principle by combining topology-preserving pore-network extraction with sequential probabilistic inference.
The pore-radius distribution is transformed into a gas-specific intra-trap step-length likelihood, $\Pi_{\mathrm{gas}}(l)$, by accounting for molecular accessibility, while the throat-length distribution supplies the likelihood of inter-trap transitions.
A Neyman--Pearson initialization followed by iterative Bayesian refinement then classifies consecutive displacements using information from both the host structure and the diffusing species.
This construction replaces repeated frame-by-frame pore matching with reusable structural likelihoods and turns matrix fluctuations from a source of geometric ambiguity into information for trajectory classification.

The trajectory simulation model (TSM) provided the controlled test that molecular dynamics trajectories cannot supply on their own.
Synthetic particles explored the extracted kerogen pore-network graph while preserving its local pore dimensions, throat lengths, connectivity, and spatial correlations, and every step retained an exact ground-truth label.
Across independently varied capture and return regimes, the benchmark exposed the characteristic failure mode of the distance-matrix (DM) classifier: rapid returns between neighboring pores were merged into apparently continuous confinement.
This error increased with the return probability and produced a substantial integral bias in the recovered capture statistics.
For $k=0.1$, $0.5$, and $0.9$, DM yielded relative deviations in $k_{\mathrm{est}}$ of $-4.0\%$, $-26.0\%$, and $-44.4\%$, respectively, whereas the corresponding deviations were $10.0\%$, $1.4\%$, and $-2.4\%$ for SIB and $7.0\%$, $1.0\%$, and $-2.6\%$ for HYB.
The high-capture regime demonstrates why a small pointwise classification error is not sufficient: class imbalance can conceal a severe distortion of the physically relevant trapping fraction.
At the same time, SIB scaled approximately as $O(N)$ with trajectory length, while DM and HYB retained the $O(N^2)$ cost of the full pairwise distance matrix.

Application to 20 methane and 20 hydrogen trajectories in type-I kerogen revealed a molecular contrast that the geometric baseline substantially distorted.
SIB and HYB produced closely agreeing trapping statistics, identifying more persistent localization for methane and more exploratory, transition-rich motion for hydrogen.
For SIB, the fitted finite-window trapping-time tail exponent was $\mu=2.381$ for methane and $\mu=3.229$ for hydrogen, while the corresponding capture-probability estimates were $k_{\mathrm{est}}=0.67$ and $0.36$.
The smaller exponent and larger capture probability for methane indicate a greater relative contribution from long trapping events, consistent with stronger confinement and molecular interactions.
Thus, incorporating the structure of the host changes the recovered physical picture of transport rather than merely improving a numerical partition of the trajectory.

The quantitative accuracy established here applies specifically to the synthetic benchmark, where trap identities and step labels are known exactly.
Agreement between SIB and HYB on molecular dynamics data supports the robustness of the inferred statistics but is not an independent ground-truth validation.
Likewise, the return probabilities used for qualitative TSM--MD comparisons were selected from trajectory morphology rather than inferred directly from the molecular dynamics trajectories.
Extending the framework will therefore require renewed tests of structural stationarity, observation-window effects, and trapping-time statistics for other materials, thermodynamic conditions, and molecular species.

Within these limits, the broader result is a transferable design principle: when local geometry is transient but its ensemble statistics are stationary, those statistics can anchor the interpretation of microscopic dynamics.
By exploiting this shift from instantaneous objects to statistical invariants, SIB converts multi-microsecond atomistic trajectories into species-resolved trapping statistics at linear computational cost.
It thereby provides a scalable path from molecular motion to the residence-time information required by reduced and continuum transport models, not only for kerogen, but for a wider class of disordered media with intrinsically dynamic internal architectures.

\section*{Data and Code Availability}
The open-source Python package implements the SIB, DM, and HYB methods for analyzing arbitrary input trajectories, including both molecular dynamics and synthetic trajectories, and also provides the TSM benchmark-trajectory generator.
It is released under the MIT License and is available as GitHub release v1.0.0 with a permanent Zenodo archive~\cite{ananev2025code}.
The molecular dynamics trajectories, structural volumes, pore-network outputs, trapping-event classifications, and derived analysis data required to reproduce the presented findings are archived separately on Zenodo under the Creative Commons Attribution 4.0 International license~\cite{ananev2025zenodo}.

\section*{ACKNOWLEDGEMENTS}
This work was supported by the Russian Science Foundation (Grant No. 25-13-00313).
We thank our colleague Dr. Irina Nesterova for her valuable feedback and support throughout the preparation of this manuscript.

\bibliographystyle{IEEEtran}
\bibliography{references}

\clearpage
\appendices
\appendix
\subsection{Comparison of trajectory analysis algorithms: computational performance}
\label{app:complexity}

In this section, we compare the computational performance of three trajectory analysis algorithms: the distance matrix (DM), structure-informed Bayesian (SIB), and hybrid (HYB) approaches.
The comparison is based exclusively on the elapsed computation time required to analyze a trajectory of a given length.

All measurements were performed on a workstation equipped with an AMD Ryzen 7 5800H processor (8 physical cores, 16 threads, base frequency up to 4.47~GHz) and 16~GB of RAM, running Ubuntu 24.04.4 LTS.
The software environment was based on Python~3.12.3.

To ensure a fair comparison, all algorithms were executed using one thread. Thus, the reported runtimes reflect the intrinsic computational cost of the methods rather than parallelization effects specific to the implementation.

\begin{figure}[h]
\includegraphics[width=0.45\textwidth]
 {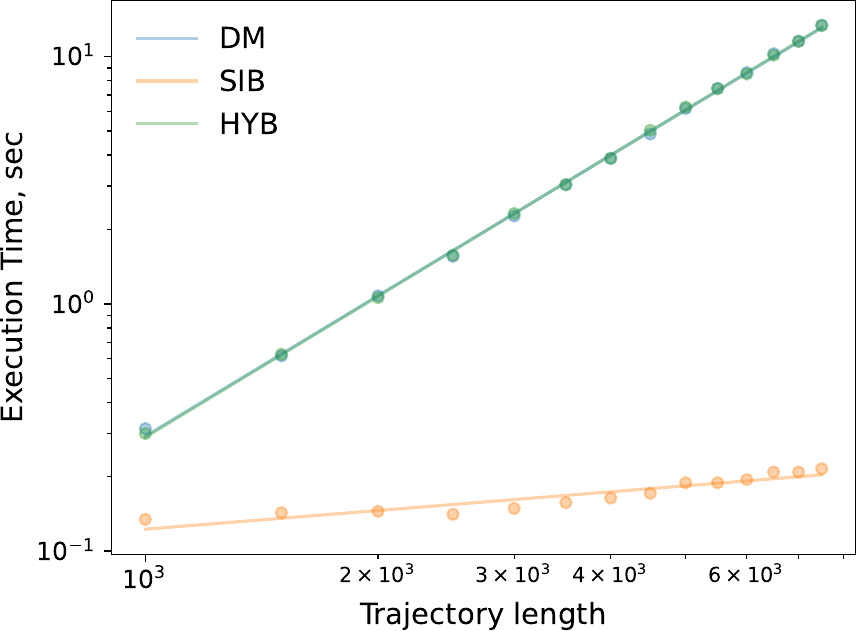}
\caption{
Computational performance of the trap identification algorithms as a function of trajectory length.
Execution time is shown versus the number of trajectory steps for DM, SIB, and HYB.
Markers indicate measured execution times averaged over multiple runs, while solid lines represent power-law fits obtained by linear regression in log--log coordinates.
}
\label{fig:complexity}
\end{figure}

For each value of $N$, an independent synthetic trajectory was generated and analyzed by each algorithm.
Reported runtimes correspond exclusively to the analysis stage and exclude trajectory generation.

For the SIB algorithm, an additional preprocessing stage is required, consisting of structural analysis of the pore space in a given kerogen cell (extraction of distributions such as $P(r)$ and $P(h)$).
However, this preprocessing is performed only once for a fixed structure and temperature and can be reused for analyzing an arbitrary number of trajectories within the same system.
Since the present comparison focuses on the analysis cost for each trajectory, the preprocessing time is not included in the reported measurements.
Figure~\ref{fig:complexity} therefore reflects only the runtime required to analyze an individual trajectory, assuming that all structural distributions are already available.

The results demonstrate distinct scaling behaviors. To quantify them, the measured runtimes were fitted in log--log coordinates using a linear regression model
\[
\log T(N) = a \log N + b ,
\]
which corresponds to the power-law dependence
\[
T(N) \approx e^{b} N^{a}.
\]
The fitted exponent \(a\) was used to characterize the effective scaling of each algorithm.

The benchmark results reveal a fundamental divergence in computational scaling that directly dictates the applicability of these frameworks to massive datasets. 
The distance matrix algorithm (DM) exhibits an effective scaling exponent of $a \approx 2.0$, directly reflecting the quadratic complexity $O(N^2)$ required to construct and process the global pairwise proximity matrix. 
This quadratic bottleneck introduces a severe computational barrier when resolving long-time transport statistics: as trajectory length $N$ increases to capture heavy-tailed trapping events, the CPU time and memory footprint of DM escalate prohibitively. 
Similarly, the hybrid (HYB) approach inherits this dominant quadratic scaling, despite utilizing the linear SIB classifier as its primary decision rule, because it still mandates the construction of the full distance matrix for the entire trajectory length.

In sharp contrast, the structure-informed Bayesian (SIB) algorithm exhibits near-perfect linear scaling with an exponent of $a \approx 1.0$.
This optimal scalability is achieved because SIB processes trajectory steps sequentially, bypassing the need for a global recurrence matrix and evaluating step-likelihoods as localized independent events. 
While SIB requires an initial structural preprocessing stage to extract the pore-network statistics, this operation is performed only once for a given matrix geometry and remains invariant over an arbitrary number of analyzed gas molecules. 
For large-scale trajectory ensembles -- such as multi-microsecond or millisecond single-particle tracking datasets generated via modern high-performance computing (HPC) platforms --the quadratic overhead of traditional geometric methods becomes completely intractable. 
Consequently, the linear computational complexity of SIB establishes it as a highly scalable, high-throughput paradigm for high-fidelity trajectory segmentation in heavily disordered media.

\subsection{Statistical assessment of temporal stationarity of pore-scale distributions}
\label{app:stationarity}

\begin{figure*}[t]
\centering
\includegraphics[width=1.\textwidth]
 {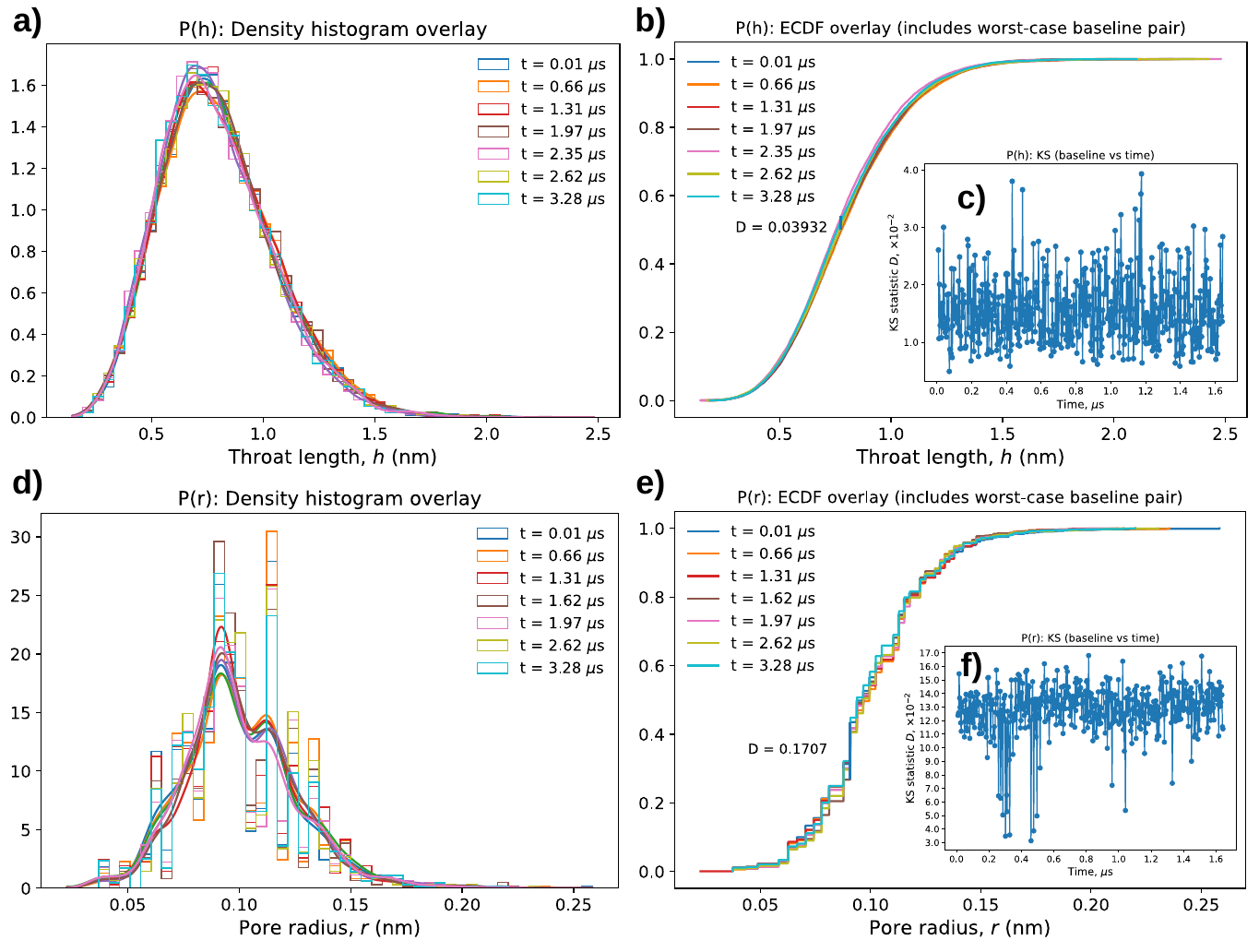}% Here is how to import EPS art
\caption{Statistical assessment of temporal stationarity of pore-scale geometric distributions.
Panels (a-c) correspond to the throat length distribution $P(h)$, while panels (d-f) show the analogous analysis for the pore radius distribution $P(r)$. Panels (a) and (d) present overlaid probability density histograms at selected time instants, illustrating the overall stability of distribution shapes. Panels (b) and (e) show the corresponding empirical cumulative distribution functions (ECDFs). The two highlighted curves represent the pair of time instants yielding the maximum Kolmogorov–Smirnov statistic, with the vertical segment indicating the KS distance $D$. Panels (c) and (f) display the evolution of the KS statistic for comparisons of the baseline with all time points, demonstrating the absence of systematic temporal drift relative to the initial configuration. In all cases, the observed KS statistics remain small and well below levels indicative of statistically significant distributional changes, confirming the effective stationarity of both $P(h)$ and $P(r)$ over the considered time interval.}
\label{fig:stat_distr}
\end{figure*}

Let $\{P_t(l)\}$ and $\{P_t(r)\}$ denote empirical samples of throat lengths $l$ and pore radii $r$, respectively, obtained at discrete time instants $t$ during the molecular dynamics simulation.
Each sample represents a one-dimensional realization drawn from an underlying continuous probability distribution $P_t(l)$ or $P_t(r)$.
Visual inspection of empirical histograms and cumulative distributions (Fig.~\ref{fig:stat_distr}) suggests only weak temporal variability of both pore-scale geometric characteristics.
To assess this observation quantitatively, we test the null hypothesis that the corresponding distributions remain invariant with respect to time.

Temporal stationarity is examined using the two-sample Kolmogorov--Smirnov (KS) test.
Given two empirical cumulative distribution functions (ECDFs) $F_n(x)$ and $G_m(x)$ constructed from samples of sizes $n$ and $m$, the KS statistic is defined as
\begin{equation}
D_{n,m} = \sup_x \left| F_n(x) - G_m(x) \right|.
\end{equation}
The null hypothesis
\begin{equation}
H_0: F(x) = G(x)
\end{equation}
states that both samples are drawn from the same underlying continuous distribution.
Under $H_0$, the asymptotic distribution of the KS statistic is given by
\begin{equation}
\Pr\!\left( \sqrt{\frac{nm}{n+m}}\, D_{n,m} \leq z \right)
= 1 - 2 \sum_{k=1}^{\infty} (-1)^{k-1} \exp\!\left(-2 k^2 z^2 \right),
\end{equation}
which allows computation of the corresponding $p$-value.
The KS test is nonparametric, does not rely on binning, and is well suited for detecting differences in continuous distributions, including those with broad or heavy-tailed behavior.

Two complementary comparison strategies were employed.
In the baseline comparison, each distribution $P_t$ was compared against a reference distribution at the initial time $t_0$.
In comparisons between consecutive time points, distributions at time instants $t$ and $t+1$ were compared in order to probe local temporal stability.
Both strategies result in multiple hypothesis tests performed within the same family, requiring correction for multiple comparisons.

Control of the familywise error rate was achieved using the Holm--Bonferroni procedure.
Given a family of $M$ hypotheses with corresponding $p$-values $\{p_i\}$ sorted in ascending order
\begin{equation}
p_{(1)} \le p_{(2)} \le \dots \le p_{(M)},
\end{equation}
the null hypothesis associated with $p_{(k)}$ is rejected if
\begin{equation}
p_{(k)} \le \frac{\alpha}{M - k + 1},
\end{equation}
where $\alpha = 0.05$ is the desired significance level.
The procedure proceeds sequentially from the smallest $p$-value and guarantees strong control of the familywise error rate while being less conservative than the classical Bonferroni correction.

The analysis was performed for a representative molecular dynamics simulation of methane ($CH_4$) in kerogen at a temperature of 300~K.
For the throat length distribution $P(h)$, neither the KS comparisons with the baseline nor those between consecutive time points revealed statistically significant differences.
All KS statistics remained small, and all hypotheses failed to be rejected after Holm--Bonferroni correction.
A summary of the maximum and median KS statistics, along with the minimum adjusted $p$-values, is reported in Table~\ref{tab:ks_stationarity}.

\begin{table}[ht]
\centering
\caption{Summary of two-sample KS test results for throat length $P(h)$ and pore radius $P(r)$ distributions for comparisons with the baseline and between consecutive time points.}
\label{tab:ks_stationarity}
\begin{tabular}{|c|c|c|c|} 
 \hline
 Test & $D_{\max}$ & $D_{\mathrm{median}}$ & $p_{\min}^{\mathrm{adj}}$ \\ [0.5ex] 
 \hline
 $P(h)$ Baseline & 0.03932 & 0.011 & 0.995 \\
 \hline
 $P(h)$ Adjacent & 0.029 & 0.012 & 0.905 \\
 \hline
 $P(r)$ Baseline & 0.1707 & 0.039 & 0.853 \\
 \hline
 $P(r)$ Adjacent & 0.062 & 0.039 & 0.876 \\ 
 \hline
\end{tabular}
\end{table}

An analogous analysis was carried out for the pore radius distribution $P(r)$.
Similarly, no statistically significant deviations were detected in either comparison strategy.
The null hypothesis of identical distributions could not be rejected in any case after correction for multiple testing, indicating the absence of measurable temporal drift in pore radii.
The detailed statistical results are summarized in Table~\ref{tab:ks_stationarity}, while representative ECDF comparisons are shown in Fig.~\ref{fig:stat_distr}b,e and the temporal evolution of the KS statistic is shown in Fig.~\ref{fig:stat_distr}c,f.

Overall, the consistently small KS statistics and uniformly large adjusted $p$-values demonstrate that both the throat length and pore radius distributions remain effectively invariant over time within the considered simulation interval.
These results quantitatively justify treating $P(h)$ and $P(r)$ as stationary distributions in the modeling framework employed in the main text.

\subsection{Neyman--Pearson Criterion and Threshold Construction}
\label{app:np_criterion}

The SIB classifier described in Section~III is implemented as two stages: an NP-based initialization that partitions trajectory steps into candidate trapped and transition sets, followed by the iterative Bayesian refinement described in Section~III.B. The label ``NP + Bayesian'' used in Fig.~\ref{fig:np_bayes_comparison} refers to this combined SIB procedure. The present appendix details the construction of the NP threshold used in the initialization step.

The initial classification of trajectory steps is based on the Neyman--Pearson (NP) lemma, which provides an optimal decision rule for testing two simple hypotheses under a constraint on the Type I error \cite{lehmann2005testing, neyman1933ix}.

Let $H_T$ denote the hypothesis that a trajectory step occurs within a trap, and $H_C$ denote the hypothesis corresponding to a transition between traps. 
Given the fitted probability densities $f_T(l)$ and $f_C(l)$, the likelihood ratio is defined as

\begin{equation}
\Lambda(l) = \frac{f_C(l)}{f_T(l)}.
\end{equation}

According to the Neyman--Pearson lemma, the most powerful test for distinguishing between $H_T$ and $H_C$ at a fixed false transition error level $\varepsilon_\mathrm{NP}$ is given by the decision rule

\begin{equation}
\Lambda(l) \mathop{\lessgtr}_{H_C}^{H_T} \eta,
\end{equation}

where $\eta$ is a threshold chosen to control the probability of Type I error.

In the present context, the Type I error corresponds to misclassifying a trapped step as a transition:

\begin{equation}
\mathbb{P}(\text{decide } H_C \mid H_T) 
= \mathbb{P}(\Lambda(l) \ge \eta \mid H_T).
\end{equation}

The threshold $\eta$ is therefore determined from the condition

\begin{equation}
\mathbb{P}(\Lambda(l)\ge\eta\mid H_T)=\varepsilon_\mathrm{NP}.
\end{equation}

In this work, $\varepsilon_\mathrm{NP}=0.01$, so the initialization labels no more than $1\%$ of trapped steps as transitions.

In the continuous setting, the threshold $\eta$ is defined implicitly by

\begin{equation}
\int_{\{l:\Lambda(l)\ge\eta\}} f_T(l)\,dl=\varepsilon_\mathrm{NP}.
\end{equation}

This condition states that the rejection region of $H_T$, where $H_C$ is selected, has probability mass $\varepsilon_\mathrm{NP}$ under $f_T(l)$.

Thus, the Neyman--Pearson threshold is determined entirely by the distribution of the likelihood ratio under the null hypothesis $H_T$.

In general, the resulting equation for the threshold does not admit an explicit analytical solution due to the nonlinear form of the likelihood ratio and the involved parametric distributions. Therefore, $\eta$ is computed numerically.

Let $\{l_j\}_{j=1}^M$ be a discretization of the support of $l$. The likelihood ratio is evaluated as

\begin{equation}
\Lambda_j = \frac{f_C(l_j)}{f_T(l_j)}.
\end{equation}

To approximate the probability measure under $f_T(l)$, weights are introduced:

\begin{equation}
w_j = \frac{f_T(l_j)}{\sum_{k=1}^{M} f_T(l_k)}.
\end{equation}

The pairs $(\Lambda_j, w_j)$ are then sorted in descending order of $\Lambda_j$, and the cumulative sum of weights is computed:

\begin{equation}
S_k = \sum_{j=1}^{k} w_{(j)}.
\end{equation}

The threshold $\eta$ is defined as the smallest value such that

\begin{equation}
S_k\ge\varepsilon_\mathrm{NP}.
\end{equation}

This procedure provides a consistent numerical approximation of the Neyman--Pearson threshold.

The resulting decision rule allows the controlled fraction $\varepsilon_\mathrm{NP}$ of trapped steps to be labelled as transitions.

In the context of trajectory analysis, this is particularly important, since incorrect identification of transitions may significantly bias the estimation of trapping statistics.

The NP-based classification thus provides an optimal initial partition of trajectory steps, which is subsequently refined using Bayesian inference.

\subsubsection{Empirical Comparison of Classification Strategies}

To assess the practical performance of the Neyman--Pearson initialization, we conducted a series of experiments on synthetic trajectories with known ground truth parameters.

\begin{figure}
\centering
\includegraphics[width=0.48\textwidth]
 {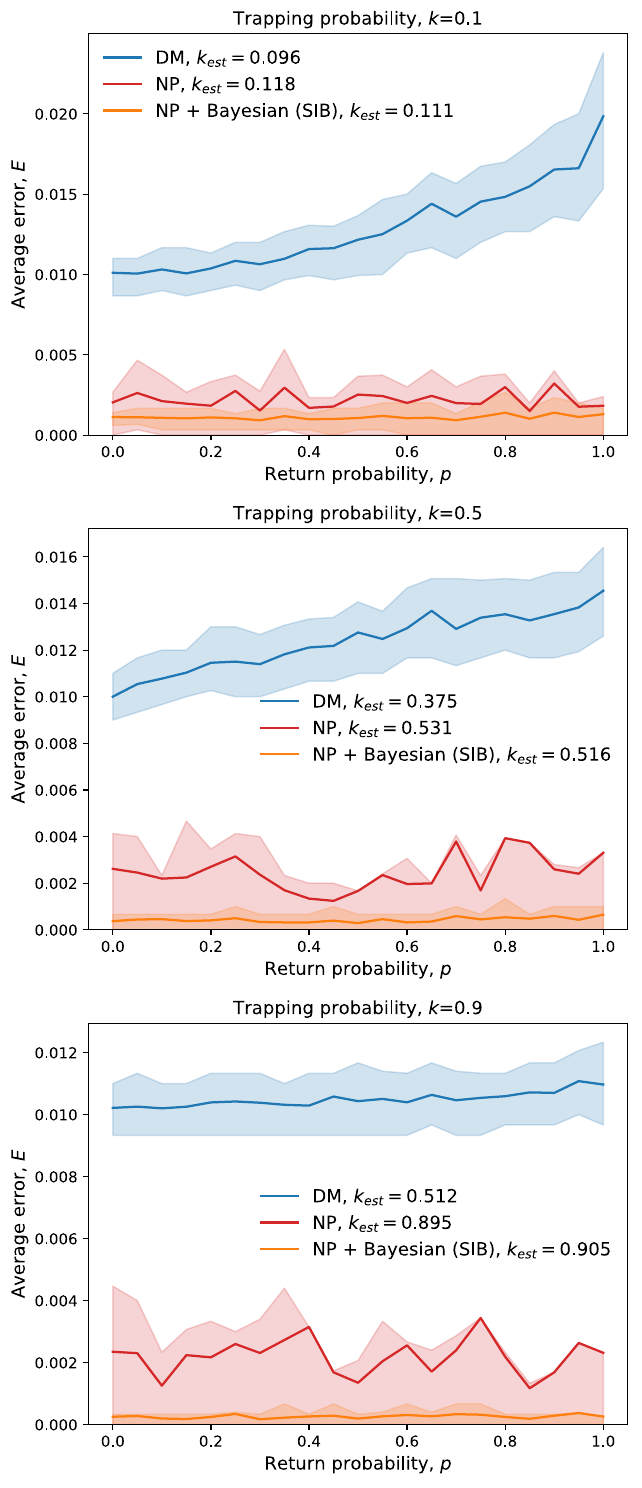}% Here is how to import EPS art
\caption{
Comparison of classification errors on synthetic trajectories for different methods: Distance matrix (DM), Neyman--Pearson (NP), and combined NP + Bayesian approach. The results are shown for three values of the trapping probability $k$ ($k=0.1$, $0.5$, and $0.9$) as a function of the return probability $p$. 
Curves show the mean error. Shaded envelopes span the 20th to 80th percentile range across realizations and include the mean where skewness places it outside that equal-tail interval.
The NP classifier exhibits a relatively wide error band, whereas the combined NP + Bayesian method significantly reduces both the average error and its variability.
}
\label{fig:np_bayes_comparison}
\end{figure}

Figure~\ref{fig:np_bayes_comparison} illustrates the average classification error as a function of the return probability $p$ for different values of the trapping probability $k$. Three approaches are compared: DM, the Neyman--Pearson (NP) classifier, and the combined NP + Bayesian refinement.

The results show that, although the NP classifier provides a theoretically optimal decision rule under a fixed Type I error constraint, its standalone application leads to a relatively wide error band and noticeable variability across the parameter range. In particular, the NP-based estimates exhibit significant dispersion, indicating sensitivity to fluctuations in the likelihood ratio.

In contrast, incorporating the Bayesian iterative refinement substantially improves the stability of the classification. The combined NP + Bayesian approach consistently reduces both the average error and its variability, yielding a significantly narrower error band across all tested regimes.

This behavior can be explained by the fact that the NP criterion operates locally on individual step likelihoods, whereas the Bayesian procedure introduces a global consistency constraint through the estimation of the prior probability $P(H_T)$. As a result, the Bayesian stage effectively regularizes the initial NP classification and suppresses spurious fluctuations.

These observations justify the use of the Neyman--Pearson classifier as an initialization step, followed by Bayesian refinement, which provides a more robust and statistically consistent classification of trajectory steps.

\subsection{Brute-Force Hyper Parameter Search for the Distance Matrix Classifier}
\label{app:dm_search}

For each candidate parameter set and each pair $(p,k)$, $E$ was averaged over 100 synthetic trajectories of 1000 points.
For each fixed $k$, these mean errors were then averaged over $p\in\{0,0.2,0.4,0.6,0.8,1.0\}$, and the parameter set with the lowest aggregate error was selected.
The resulting set was kept fixed for all $p$ at that $k$.
This procedure tests robustness to different degrees of cyclic trapping, avoids fitting parameters to an individual $p$ or trajectory, and provides one transferable parameter set for each trapping probability.

The twelve candidate scale sets were the full set $\{0.5,1.0,1.5,2.0,2.5,3.0\}$, the five subsets $\{1.5,2.0,2.5\}$, $\{0.5,1.0\}$, $\{2.5,3.0\}$, $\{0.5,3.0\}$, and $\{1.5,2.0\}$, and the six single scale sets $\{\lambda\}$ with $\lambda\in\{0.5,1.0,1.5,2.0,2.5,3.0\}$.
For each scale set, the search included the reference motion type, Brownian motion (Bm) or fractional Brownian motion (fBm), $v_c\in\{0.1,0.5,0.9\}$, smoothing window width $\mu\in\{1,3,5,7\}$, diagonal percentile in $\{0,10,50\}$, and $p_\mathrm{val}\in\{0.01,0.1,0.9\}$.
The diagonal percentile determines the number $s$ of adjacent diagonals filled in the recurrence matrix, and the optimal search selected $s=0$ for all three trapping probabilities $k$.
The different numbers of scales in Table~\ref{tab:dm_params} are therefore outcomes of the common search rather than manual choices after optimization.
The selected configuration uses fractional Brownian motion (fBm) as the reference motion for $k=0.1$ and $k=0.5$, and Brownian motion (Bm) for $k=0.9$.

Table~\ref{tab:dm_params} reports the selected parameter sets.
Here, $\{\lambda_i\}$ is the scale set, $\mu$ is the smoothing window width, $v_c$ is the block-invariant threshold, and $p_\mathrm{val}$ is the statistical threshold parameter defined in Section~III.A.

\begin{table}[ht]
\centering
\caption{ Optimized DM parameter sets used by DM and HYB.}
\label{tab:dm_params}
\begin{tabular}{|c|c|c|c|c|}
\hline
$k$ & $\{\lambda_i\}$ & $\mu$ & $v_c$ & $p_\mathrm{val}$ \\
\hline
$0.1$ & $0.5,\;1.0$ & $3$ & $0.5$ & $0.1$ \\
\hline
$0.5$ & $1.5,\;2.0,\;2.5$ & $3$ & $0.5$ & $0.9$ \\
\hline
$0.9$ & $0.5,\;1.0,\;1.5,\;2.0,\;2.5,\;3.0$ & $1$ & $0.1$ & $0.9$ \\
\hline
\end{tabular}
\end{table}

\end{document}